\documentclass[preprint,12pt]{elsarticle}

\usepackage{amssymb}
\usepackage{amsmath}
\usepackage{setspace}
\usepackage{moreverb}
\usepackage{graphicx}
\usepackage{epstopdf}
\usepackage{amsfonts}
\usepackage{array}
\usepackage{subfig}
\usepackage{caption}

\journal{Physical Communication}

\begin{document}

\begin{frontmatter}



\title{Unsupervised Learning Approach for Beamforming in Cell-Free Integrated Sensing and Communication}


\author{Mohamed Elrashidy} 

\affiliation{organization={Electrical Engineering Department, King Fahd University of Petroleum and Minerals},
            city={Dhahran},
            country={Saudi Arabia}}

\author{Mudassir Masood} 

\affiliation{organization={Electrical Engineering Department, Center for Communication Systems and Sensing, King Fahd University of Petroleum and Minerals},
            city={Dhahran},
            country={Saudi Arabia}}

\author{Ali Arshad Nasir} 

\affiliation{organization={Electrical Engineering Department, Center for Communication Systems and Sensing, King Fahd University of Petroleum and Minerals},
            city={Dhahran},
            country={Saudi Arabia}}

\begin{abstract}
Cell-free massive multiple input multiple output (MIMO) systems  can provide reliable connectivity and increase user throughput and spectral efficiency of integrated sensing and communication (ISAC) systems. This can only be achieved through intelligent beamforming design. While many works have proposed optimization methods to design beamformers for cell-free systems, the underlying algorithms are computationally complex and potentially increase fronthaul link loads. To address this concern, we propose an unsupervised learning algorithm to jointly design the communication and sensing beamformers for cell-free ISAC system. Specifically, we adopt a teacher-student training model to guarantee a balanced maximization of sensing signal to noise ratio (SSNR) and signal to interference plus noise ratio (SINR), which represent the sensing and communication metrics, respectively. The proposed scheme is decentralized, which can reduce the load on the central processing unit (CPU) and the required fronthaul links. To avoid the tradeoff problem between sensing and communication counterparts of the cell-free system, we first train two identical models (teacher models) each biased towards one of the two tasks. A third identical model (a student model) is trained based on the maximum sensing and communication performance information obtained by the teacher models. While the results show that our proposed unsupervised DL approach yields a performance close to the state-of-the-art solution, the proposed approach is more computationally efficient than the state of the art by at least three orders of magnitude.
\end{abstract}


\begin{keyword}


cell-free \sep integrated sensing and communication \sep beamforming \sep unsupervised learning \sep deep learning \sep teacher-student

\end{keyword}

\end{frontmatter}



\section{Introduction}
\label{sec:intro}

Integrated sensing and communication (ISAC) is a promising 6G technology that is expected to support many applications such as radar \cite{dehkordi2023active}, localization \cite{gao2023cooperative}, vehicular networks \cite{liu2022deep}, \cite{yuan2021integrated}, \cite{cheng2022integrated}, and UAV assisted networks \cite{meng2023uav}. However, the performance of ISAC systems is affected by a tradeoff between communication quality and sensing accuracy. Dong \textit{et al.} \cite{dong2022sensing} proposed a general scheme for resource allocation in ISAC systems and studied the mentioned tradeoff between target detection and communication quality of service (QoS). While some works focused on structured beamforming and optimized power and bandwidth allocation coefficients in different ISAC MIMO systems \cite{dong2022sensing,dong2022localization,zeng2023integrated}, it can be argued that optimizing unstructured beamformers is more general since the allocated power coefficients are readily included in the optimized beam vectors. Xu \textit{et al.} \cite{xu2022robust} formulated a beamforming problem for ISAC systems to maximize the sum secrecy rate while maintaining a certain user QoS. After handling the non-convexity of the problem, the authors proposed an iterative algorithm based on block coordinate descent to obtain a suboptimal solution. He \textit{et al.} \cite{he2023full} proposed two problems to design the transmit and receive beamformers and signal covariance matrix through minimizing the transmit power or maximizing the sum rate in ISAC systems. The two problems are subject to sensing and communication quality thresholds. Hua \textit{et al.} \cite{hua2021transmit} proposed an ISAC downlink design by minimizing the beam pattern matching error, where pre-designed beam patterns are considered. The minimization is subjected to a certain signal-to-interference plus noise ratio (SINR) threshold, and the optimization parameters are the beamformers and radar signal covariance matrix. Many other works addressed the beamforming problem in reconfigurable intelligent surface (RIS) assisted ISAC systems \cite{yu2022location,liu2022joint,yu2023active,zhong2023joint}.

The use of deep learning (DL) to design the downlink of general ISAC systems has recently been explored \cite{demirhan2023integrated,salem2023data}. Particularly, Liu \textit{et al.} \cite{liu2022deepirs} trained a couple of deep neural networks (DNNs) to estimate sensing and communication channels for an ISAC system. Liu \textit{et al.} \cite{liu2022deep} used a recurrent neural network (RNN) that is composed of convolutional long-short term memory (CLSTM) subunits. Such network is expected to capture the spatial and temporal properties of the environment. The network was trained to predict the angles of the vehicles within the range of an ISAC multi-antenna BS. The predicted angles were used for designing a predictive transmit beamformer. The training was done offline in a supervised fashion. A similar approach is adopted by Liu \textit{et al.} \cite{liu2022learning}, where an RNN consisting of CLSTM subunits was used for direct beamforming in the same system considered by \cite{liu2022deep}. In \cite{liu2022learning}, the authors first proposed a sum-rate maximization problem constrained by Cramer-Rao lower bound as a sensing error metric. Solving this problem provided means to both generate an unsupervised dataset and a baseline for comparison between the conventional CVX solver and the unsupervised learning scheme. Mu \textit{et al.} \cite{mu2021integrated} proposed an FCNN that predicts the beamformer and the angular parameters of the detected vehicles given the received signal at the ISAC-assisted roadside units. Liu \textit{et al.} \cite{liu2023distributed} proposed a DL network architecture to manage interference in ISAC systems through power allocation design. To that end, the authors utilized unsupervised learning for training. They also used transfer learning to predict a proper beamforming scheme for interference management. 


The aforementioned works considered single cell systems where beamformers are designed at a single base station (BS) to serve a number of users within a designated area. Cell-free systems, on the other hand, are promising in terms of meeting the demands of next generation technologies, such as energy and spectral efficiency, low latency and reliability \cite{elhoushy2021cell}. It was found that cell-free schemes outperform conventional small-cell networks tremendously in terms of 95\%-likely per-user throughput, energy efficiency and spectral efficiency \cite{nayebi2017precoding,bjornson2019making,ngo2017cell,ngo2020correction}. Nevertheless, only few recent works considered beamforming design problems in cell-free ISAC systems. Demirhan and Alkhateeb \cite{demirhan2023cell} proposed a joint sensing and communication beamforming optimization problem, where the objective is to jointly maximize sensing and communication metrics, namely, sensing signal-to-noise ratio (SSNR) and SINR, respectively, subjected to the power budget constraint at each access point (AP). Their results show that the joint optimization problem outperforms other methods where sensing or communication beams are designed a priori. Liu \textit{et al.} \cite{liu2023joint} considered the problem of AP selection (i.e., transmitting or receiving modes) to improve the degrees of freedom in cell-free ISAC networks. The authors formulated a sum-maximization problem to jointly optimize the AP selection mode, the beamformer vectors and the receive filter vectors. Other works that studied the beamforming problem in cell-free ISAC networks include \cite{behdad2023multi,mao2023communication,elfiatoure2023cell,cao2023joint}.

Since studies on ISAC-assisted cell-free MIMO systems are limited and recent, the application of DL techniques for solving power allocation or beamforming design has not been considered thus far. The only exception is the recent work in \cite{zeng2023integrated}, where the authors considered a deep reinforcement learning (DRL) solution to optimize power coefficients for the data and the pilot in cell-free ISAC systems. However, this solution requires discretizing the optimization variables, since the action space is usually discrete. Also, the work did not consider solving a more general and challenging beamforming design problem.

It is worth mentioning that many works utilized DL solutions to allocate resources in cell-free MIMO and massive MIMO wireless non-dual systems (i.e., where ISAC is not deployed) \cite{iliadis2022road}. The authors of \cite{zhao2022power}, for example, addressed three power allocation problems in cell-free massive MIMO systems through deep supervised learning and DRL. Zaher \textit{et al.} \cite{zaher2022learning} formulated a centralized optimization problem for power allocation in cell-free networks. The authors proposed a distributed DNN solution highlighting one advantage of using DL techniques -compared to the conventional solvers for the formulated problem- by alleviating the need of using fronthaul links for collecting information from different APs. While \cite{zaher2022learning} considered a supervised training scheme, where the problem is first solved through conventional CVX-based solutions to generate a dataset, many works considered unsupervised learning for cell-free massive MIMO systems \cite{nikbakht2019unsupervised,hojatian2022decentralized,zhang2023unsupervised,rajapaksha2023unsupervised}. Nikbakht \textit{et al.} \cite{nikbakht2019unsupervised} reduced the dimension of the input to the DNN by using the effective gain vector instead of the concatenated large scale coefficients, which enabled them testing the performance on massive MIMO systems with large number of APs. However, such solution is not possible without collecting information from all APs at a central processing unit (CPU). In \cite{hojatian2022decentralized}, two unsupervised training paradigms were studied, a decentralized paradigm (i.e., distributed) and a partially decentralized paradigm, to perform hybrid beamforming in free-cell massive MIMO systems. Zhang \textit{et al.} \cite{zhang2023unsupervised} proposed unsupervised training schemes to solve the power allocation problem for the uplink and the downlink of cell-free massive MIMO systems. Rajapaksha \textit{et al.} \cite{rajapaksha2023unsupervised} considered hardware impairment in their problem design. Other works focusing on employing DL (supervised or unsupervised) or DRL for resource allocation in MIMO or massive MIMO cell-free systems include \cite{rajapaksha2021deep,zhang2021deep,luo2022downlink}. The mentioned works focused on traditional wireless communication problems only, which is profoundly simpler than ISAC systems.


The joint communication and sensing beamforming design by leveraging computationally efficient unsupervised learning approach is an open research problem, and is the focus of our work. The challenges entailed by this problem mainly include the high dimensionality of the prediction beamforming vectors and balancing the sensing and communication performance considering the tradeoff between the two tasks. Inspired by the teacher-student architectures used in recent DL works for object detection tasks (e.g., \cite{liu2021unbiased,vandeghen2022semi}), we train two identical teacher models to obtain the maximum expected performance of the underlying model in terms of sensing alone and communication alone. This information (i.e., the maximum expected performance of the model) helps controlling the bias of the loss function that is used to train a student model, forcing the model to achieve the required balance between sensing and communication. As such, for models with sufficient complexity, the proposed method generates beamformers with as high quality as the state-of-the-art methods but with a much better time complexity, which is crucial for real-time and delay-sensitive ISAC systems.

To that end, the contributions of this work are summarized as follows.
\begin{enumerate}
    \item A novel DL approach to solve the joint beamforming problem in cell-free massive MIMO ISAC systems is proposed. To the best of our knowledge, a DL solution for this particularly promising setting has not been pursued in the existing literature

    \item The proposed DL training scheme is unsupervised. As such, the resources consumed for generating supervised dataset -including the time required for solving the problem numerously using computationally complex solutions- are substantially reduced.

    \item Rather than training a single DNN that predicts the set of beamformers for all APs, we jointly train a set of DNNs, each of which predicts the beamformer of a single AP. Such distributed approach enables a fundamental reduction in fronthaul link load compared to centralized approaches. To achieve the distribution aspect of the proposed DL method, cooperation between the DNNs only occurs during calculation of the loss, which is only needed during the offline training phase. This way, every DNN can generate the beamformers of its corresponding AP during real-time deployment independently (i.e., without cooperation with the other APs).

    \item A novel form of a teacher-student architecture is utilized to prevent biases towards one functionality of ISAC systems (i.e., either sensing or communication). Instead of manually tuning a hyperparameter in the loss function to seek the target balance between sensing and communication, two teacher models are trained so that each teacher is completely biased towards one of the two functionalities of ISAC. Given the performances of these biased teachers, a third model, namely the student model, can learn how to balance the two functionalities through the training phase.


    \item The proposed DL approach generates high quality beamformers and yields a profoundly much higher computational efficiency compared to the existing CVX-solver based approach.

\end{enumerate}

The rest of the paper is organized as follows. System model is described in Section \ref{sec:sys}. The joint communication and sensing optimization problem is discussed in Section \ref{sec:optim}. The unsupervised losses used for training the teachers and the student are stated in Section \ref{sec:loss}. The training paradigm details are described in Section \ref{sec:training}. Model architectures at which the proposed methodology is evaluated are described in Section \ref{sec:archs}. Results and discussion are in Section \ref{sec:results}, then the work is concluded in Section \ref{sec:conc}.

\textbf{Notation}: We use boldface for vectors and matrices, where vectors are represented by lowercase letters ($\mathbf{v}$) and matrices are represented by uppercase letters ($\mathbf{V}$). We define $\mathbf{I}_N$ as the identity vector of a size $N\times N$. We use a format for sets such as $\mathcal{A},\mathcal{B}$ and $\mathcal{C}$. $\mathbb{R}$ and $\mathbb{C}$ represent the sets of real and complex numbers, respectively.The notation $\mathcal{C N}(m,\mathbf{C})$ represents the complex Gaussian distribution with a mean, $m$, and a covariance matrix, $\mathbf{C}$. The operator $\operatorname{blk}\left[\mathbf{A}_1, \mathbf{A}_2 \right]$ stacks the 2D matrices, $\mathbf{A}_1$ and $\mathbf{A}_2$, along a third dimension. The operators $\mathfrak{R}(\cdot)$ and $\mathfrak{I}(\cdot)$ represent the real and imaginary parts of a complex tensor.


\section{System Model}
\label{sec:sys}

\begin{figure}[t]
     \centering
     \includegraphics[width=\linewidth]{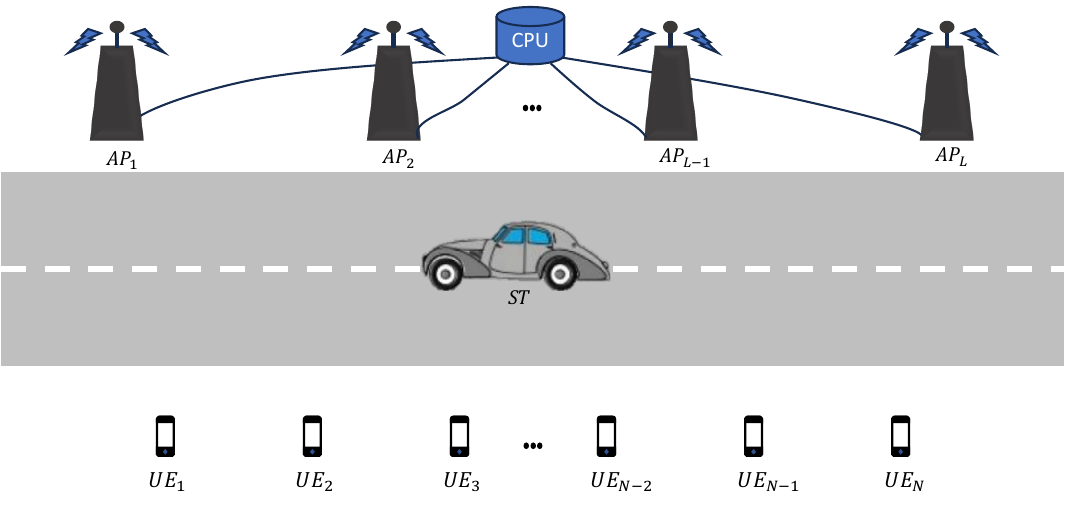}
     \caption{Cell-free ISAC example setup.}
     \label{fig:isacSetup}
\end{figure}

Consider a cell-free ISAC system with $L$ APs, $N$ UEs and a single sensing target (ST). Every AP is equipped with $M$ antennas and is capable of transmission and reception simultaneously. All APs are connected to a CPU through fronthaul links for administration and synchronization. Figure \ref{fig:isacSetup} depicts an example of the described system. Let the AP index set be denoted by $\mathcal{L}$. Let $\mathcal{Q}\triangleq\{1,...,N+1\}$ be the index set of the stacked beamforming vectors for all users and the target, $\{\mathbf{w}_q\}_{q\in\mathcal{Q}}$, where $\mathbf{w}_q\triangleq[\mathbf{w}_{1q}^{\top},...,\mathbf{w}_{Lq}^{\top}]^{\top}$ is the stacked $q$-th beam from all APs and $\mathbf{w}_{lq}\in\mathbb{C}^{M\times1}$ is the beam dedicated by the $l$-th AP to the $q$-th agent (i.e., a UE or an ST). Define the subsets $\mathcal{N}\triangleq\{1,...,N\}$ and $\mathcal{S}\triangleq\{N+1\}$ to denote the stacked communication beam indices and the stacked sensing beam index in the set $\mathcal{Q}$, respectively. The first $N$ beam indices in $\mathcal{Q}$ are reserved for the UEs, whereas the last index of $\mathcal{Q}$ is reserved for the ST. Moreover, define the stacked communication channel vector for the $n$-th UE as $\mathbf{h}_n\triangleq[\mathbf{h}_{1n}^{\top},...,\mathbf{h}_{Ln}^{\top}]^{\top}$, where $n\in\mathcal{N}$ and $\mathbf{h}_{ln}\in\mathbb{C}^{M\times1}$. 

User assignment is not considered in this model, meaning that all APs serve all users and sense the targets. Consequently, the received signal at the $n$-th UE is given by

\begin{equation}
\label{eq:receivedISAC}
y_n = \sum_{q\in\mathcal{Q}} \mathbf{h}_n^H\mathbf{w}_qx_q+n_n,
\end{equation}
where $n_n \sim \mathcal{C N}\left(0, \sigma_n^2\right)$ is the UE's receiver noise and $x_q$ is the symbol sent for the $q$-th agent. Notice that all APs transmit the same message to each agent, which explains why $x_q$ is not a function of the AP index, $l$.

The expression in (\ref{eq:receivedISAC}) is composed of the desired signal at $q=n$, communication interference at $q\in\mathcal{N} \backslash\{n\}$, sensing interference at $q\in\mathcal{S}$, and noise. Assuming a unit power symbol-based design (i.e., $\mathbb{E}\left[|x_q|^2\right]=1 ~\forall q$), the SINR at which the $n$-th UE decodes the received signal is given by

\begin{equation}
\label{eq:SINR}
\operatorname{SINR}_n=\frac{\left|\mathbf{h}_n^H \mathbf{w}_n\right|^2}{\sum_{n^{\prime} \in \mathcal{N} \backslash\{n\}}\left|\mathbf{h}_n^H \mathbf{w}_{n^{\prime}}\right|^2+\sum_{s \in \mathcal{S}}\left|\mathbf{h}_n^H \mathbf{w}_s\right|^2+\sigma_n^2}
\end{equation}

While characterizing the communication metric, SINR, requires signal analysis at the user's end, characterizing the sensing metric requires signal analysis at the receiver AP. The received signal at the $r$-th AP is given by

\begin{equation}
\label{eq:receivedISAC2}
\mathbf{y}_r=\sum_{l \in \mathcal{L}} \alpha_{lr} \mathbf{a}\left(\theta_{r}\right) \mathbf{a}^H\left(\theta_{l}\right) \mathbf{x}_{l}+\mathbf{n}_{r},
\end{equation} 
where $\alpha_{lr} \sim \mathcal{C N}(0,\sigma_{s_{lr}}^2)$ is the effective sensing channel gain, $\mathbf{a}\left(\theta_{r}\right)\in\mathbb{C}^{M\times1}$ is the steering vector given by

\begin{equation}
\label{eq:steering}
\mathbf{a}\left(\phi\right) = [1,e^{j2\pi\lambda cos(\phi)},...,e^{j2\pi\lambda (M-1)cos(\phi)}],
\end{equation} 
where $\lambda$ is the wavelength-antenna spacing ratio, $\theta_{l}$ is the angle of departure from the $l$-th AP, $\theta_{r}$ is the angle of arrival at the $r$-th AP, $\phi\in\{\theta_{l},\theta_{r}\}$, $\mathbf{n}_{r}\sim \mathcal{C N}(0,\sigma_{a_{r}}^2\mathbf{I}_M)$ is the receiver antenna noise vector at the $r$-th AP, and $\mathbf{x}_{l}\in\mathbb{C}^{M\times1}$ is the transmitted signal from the $l$-th AP which is defined as

\begin{equation}
\label{eq:transmitISAC}
\mathbf{x}_{l} = \sum_{q\in\mathcal{Q}} \mathbf{w}_{lq}x_q.
\end{equation}

To assist the sensing quality, we consider the SSNR formula defined as follows

\begin{equation}
\label{eq:SSNR}
\mathrm{SSNR}=\frac{\sum_{r \in\mathcal{L}} \sum_{l \in \mathcal{L}} \sigma_{s_{lr}}^2\left\|\mathbf{a}^H\left(\theta_{l}\right) \overline{\mathbf{W}}_{l}\right\|^2}{\sum_{r \in \mathcal{L}} \sigma_{a_{r}}^2},
\end{equation}
where $\overline{\mathbf{W}}_{l}\triangleq[\mathbf{w}_{l1},...,\mathbf{w}_{l(N+1)}]$. Notice that all beams contribute to sensing rather than considering them as interference. This can be justified by the fact that the sensing functionality requires analyzing the incident signal to track its reflection pattern and does not require decoding the information content within the received signal.


\section{Joint Communication and Sensing Beamforming Problem}
\label{sec:optim}

The joint cell-free ISAC optimization problem is given by

\begin{subequations}
\label{eq:jsc}
\begin{flalign}
\max_{\{\mathbf{w}_{lq}\}_{q\in\mathcal{Q}} ~ \forall l} \{g_1,g_2\} \\ &
\text { s.t. } \sum_{q \in \mathcal{Q}}\left\|\mathbf{w}_{lq}\right\|^2 \leq P_l, \quad \forall l \in \mathcal{L},
\end{flalign}
\end{subequations}
where $g_1 \triangleq \mathrm{SSNR}$, $g_2\triangleq \min \{\mathrm{SINR}_n\}_{n\in\mathcal{N}}$ and $P_l$ is the power budget at the $l$-th AP. 

Following \cite{demirhan2023cell}, problem (\ref{eq:jsc}) is addressed by maximizing SSNR while introducing a constraint for the SINR, where the threshold of the SINR constraint is evaluated by finding a feasible solution for the following problem

\begin{equation}
\label{eq:optPriority}
\begin{array}{ll}
\text { find } & \{\mathbf{w}_{ln}\}_{n\in\mathcal{N}} ~ \forall l \\
\text { s.t. } & \operatorname{SINR}_n \geq \gamma, \quad \forall n \in \mathcal{N} \\
& \sum_{n \in \mathcal{N}}\left\|\mathbf{w}_{l n}\right\|^2 \leq \rho P_l, \quad \forall l \in \mathcal{L},
\end{array}
\end{equation} where $0\leq\rho\leq 1$ is the communication power ratio, which dedicates a portion of the power budget to communication beams. The sensing beam is assumed to be in the null-space of the communication channels to compute the SINR. Problem (\ref{eq:optPriority}) is solved repeatedly for a certain range of $\gamma$. The objective is to find the highest value of $\gamma$ at which problem (\ref{eq:optPriority}) is feasible. As this value, denoted by $\gamma_{high}$, is obtained, problem (\ref{eq:jsc}) can be re-written as

\begin{subequations}
\label{eq:jsc2}
\begin{flalign}
\max_{\{\mathbf{w}_{lq}\}_{q\in\mathcal{Q}} ~ \forall l} g_1 \\ &
\text { s.t. } g_2 \geq \gamma_{high} \\ &
\sum_{q \in \mathcal{Q}}\left\|\mathbf{w}_{lq}\right\|^2 \leq P_l, \quad \forall l \in \mathcal{L}.
\end{flalign}
\end{subequations}

As such, the mentioned joint optimization strategy requires solving two optimization problems. The first optimization problem, (\ref{eq:optPriority}), requires applying bisection search for $\gamma$. The second optimization problem, (\ref{eq:jsc2}), requires semidefinite programming (SDP) \footnote{Addressing the convexity of problems (\ref{eq:optPriority}) and (\ref{eq:jsc2}) is not within the scope of this work. Instead, the reader is referred to \cite{demirhan2023cell} for such details.}. Such complex solution is impractical for the system depicted in Figure \ref{fig:isacSetup}, since the sensing target is not moving slowly. To address this complexity concern, we introduce an unsupervised DL-based solution to problem (\ref{eq:jsc}) that can be deployed in real time.

\begin{figure}[t]
     \centering
     \includegraphics[width=0.9\linewidth]{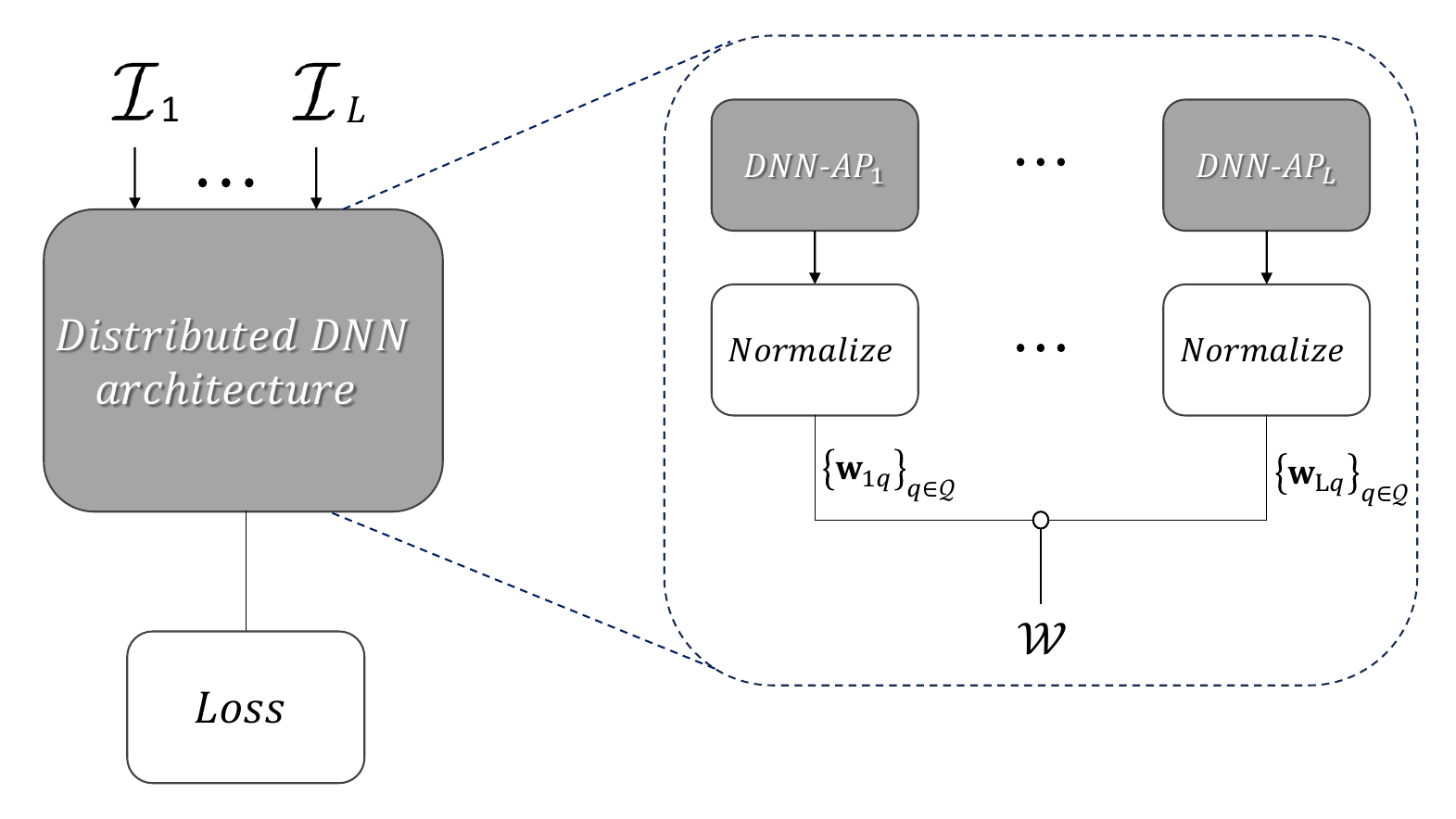}
     \caption{The proposed training unit for cell-free ISAC beamforming. The shaded boxes are trainable parameters.}
     \label{fig:trainingUnit}
\end{figure}


\section{The Unsupervised Loss}
\label{sec:loss}

Refer to the unsupervised training scheme depicted in Figure \ref{fig:trainingUnit}. We train $L$ DNNs so that the $l$-th DNN evaluates the $l$-th AP's beamforming vectors, $\{\mathbf{w}_{lq}\}_{q\in\mathcal{Q}}$, independently (i.e., without requiring information from the other APs). Thus, the proposed training scheme is distributed. Also, we denote the input of the $l$-th DNN by $\mathcal{I}_l$. Detailed description of the training unit in Figure \ref{fig:trainingUnit} and its relation to the teacher-student technique is found in Section \ref{sec:training}. The DNN architectures and the input-output expressions are detailed in Section \ref{sec:archs}. This section's main focus is explaining the loss block in Figure \ref{fig:trainingUnit}.

Different losses are used to train the teacher and student models. Given the concatenated prediction, $\mathcal{W}$ (c.f., Figure \ref{fig:trainingUnit}), two unsupervised losses are utilized to evaluate and update the $L$ DNNs. The first loss is called the teacher loss, and is given by

\begin{equation}
\label{eq:teacher_unsupLoss}
    L_t^{(\beta)} = -[(1-\beta)g_1 + \beta g_2],
\end{equation} where $0\leq\beta\leq 1$ is a normalization parameter that controls the balance between $g_1$ and $g_2$. Specifically, when $\beta \rightarrow 0$, the model tends to maximize $g_1$ and ignores $g_2$. In other words, the model becomes biased towards sensing. On the other hand, when $\beta \rightarrow 1$, the model tends to maximize $g_2$ and ignores $g_1$, preferring communication over sensing. Since fine-tuning $\beta$ so that the model equally focuses on maximizing both SSNR and SINR is difficult, we exploit the described behavior of the teacher loss to train two biased models. One model is biased towards sensing, namely the SSNR teacher, and the other model is biased towards communication, namely the SINR teacher.

The second loss is called the student loss, which is defined as follows
\begin{equation}
\label{eq:student_unsupLoss}
    L_s^{(\lambda)} = -\left[(1-\lambda) \frac{g_1}{\mathbb{E}\left[g_1^{(max)}\right]} + \lambda \frac{g_2}{\mathbb{E}\left[g_2^{(max)}\right]}\right],
\end{equation} where $g_1^{(max)}$ and $g_2^{(max)}$ are respectively the maximum possible SSNR and the maximum possible minimum SINR values that can be attained by a certain model architecture, and $0\leq\lambda\leq 1$ is a normalization parameter that is adapted according to the gap between the prediction scores (i.e., $g_1$ and $g_2$) and the maximum attainable scores (i.e., $g_1^{(max)}$ and $g_2^{(max)}$).

While the behavior of the student loss with respect to $\lambda$ is similar to the teacher loss behavior with respect to $\beta$, as detailed above, there is a difference between the two losses. That is, the student loss is bounded between -1 and 0 unlike the teacher loss, whose lower bound is dependent on the SSNR and SINR.

The maximum attainable scores, $g_1^{(max)}$ and $g_2^{(max)}$, are obtained from the trained SSNR teacher and the SINR teacher, respectively. Instead of evaluating $g_1^{(max)}$ and $g_2^{(max)}$ at every iteration, the trained teacher models are evaluated on the training set once before the student training, and the mean value of the teacher scores is taken, hence the expected value operator in the denominator of (\ref{eq:student_unsupLoss}).


\section{Training Scheme}
\label{sec:training}

\subsection{Training units and distributed models}
\label{subsec:distrib}


We propose a distributed unsupervised training unit that jointly maximizes $g_1$ and $g_2$. Specifically, $L$ DNNs are trained to predict suitable beamforming vectors for the $L$ APs. The generated $L$ predictions are normalized, concatenated, and fed to the loss function as shown in Figure \ref{fig:trainingUnit}. The evaluated loss is used to update the parameters of the $L$ DNNs through backpropagation. Notice that the training loss is the only way through which the different DNNs cooperate. After the training phase, which is done offline, every DNN evaluates its own beamforming vectors without cooperation with the other DNNs. Notice from Figure \ref{fig:trainingUnit} that a training unit can feature different losses through the $Loss$ block. For simplification, the input to a training unit block is not only the initialized distributed DNN weights but also the loss used throughout the training, whereas the output of a training unit block is the trained distributed DNNs.

The input to the $l$-th model is the channel state information (CSI) of the $l$-th AP. The exact form of the input CSI depends on the DNN architecture considered, hence it is discussed alongside the architectures in Section \ref{sec:archs}. CSI is assumed to be available at each AP through existing centralized or decentralized estimation methods \cite{xu2021privacy,song2021joint,amadid2022channel}. Notice that CSI consists of both communication and sensing channel information. The output of the $l$-th network is the set of beamforming vectors for all agents, $\{\mathbf{w}_{lq}\}_{q\in\mathcal{Q}} ~ \forall l$. The output of every network is normalized to guarantee the power constraints mentioned in (\ref{eq:jsc2}c). Predicting $\{\mathbf{w}_{lq}\}_{q\in\mathcal{Q}}$ given the CSI of the $l$-th AP enables distributing the trained $L$ networks among the APs.

\subsection{Teacher-Student scheme}
\label{subsec:teach_Stu}


To avoid looking for suitable $\beta$ in (\ref{eq:teacher_unsupLoss}) that achieves a balanced maximization for $g_1$ and $g_2$ through a time-consuming grid search, a teacher-student scheme is adapted instead. Specifically, a distributed model is trained to generate beamformers that solely maximize the sensing performance with complete disregard of the communication aspect of the system. This is achieved by applying the aforementioned training unit with a loss function, $L_t^{(\beta=0)}$, where $L_t^{(\beta)}$ is defined by (\ref{eq:teacher_unsupLoss}). Contrarily, another distributed model is trained to focus on maximizing the communication performance. The training unit used in this case is characterized by the loss, $L_t^{(\beta=1)}$. Ultimately, neither of the two distributed models can be used for a proper ISAC beamforming design, since each distributed model is biased towards either sensing or communication. Nevertheless, these two distributed models can provide information about the maximum sensing and communication performance that can ever be attained by the specific DNN architecture. As such, these two pieces of information are used to guide the training of a third distributed model, where the used training block is characterized by $L_s^{(\lambda)}$. This way, the third distributed model can learn the proper balance between sensing and communication.

\begin{figure}[t]
     \centering
     \includegraphics[width=\linewidth]{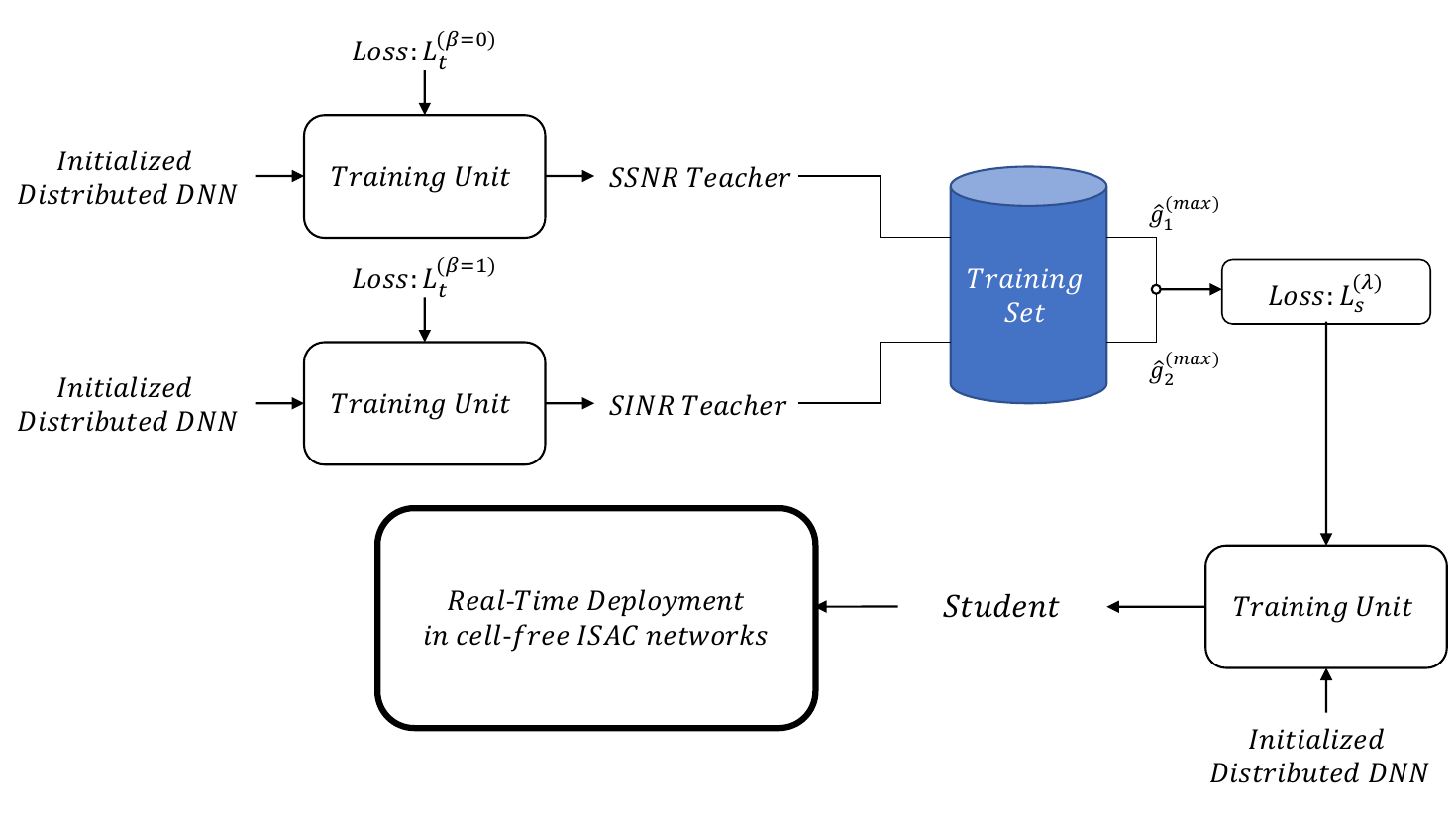}
     \caption{The proposed teacher-student training paradigm.}
     \label{fig:teach_stu}
\end{figure}

For simplicity, the sensing-biased distributed model is called the SSNR teacher model, the communication-biased distributed model is called the SINR teacher model, and the balanced distributed model is the student model. The three models are identical in structure. The maximum performance information of sensing and communication is extracted by applying the trained teacher models to the training set. The predicted beamformers corresponding to the training points are assisted by evaluating $g_1$ and $g_2$ for every sample. We consider the sample means of the scores, $g_1$ and $g_2$, as sufficient indicators of the maximum performance information needed to train the student using the loss in (\ref{eq:student_unsupLoss}). The expectation operators in (\ref{eq:student_unsupLoss}) are approximated as $\mathbb{E}\left[g_1^{(max)}\right] \approx \hat{g}_1^{(max)}$ and $\mathbb{E}\left[g_2^{(max)}\right] \approx \hat{g}_2^{(max)}$, where $\hat{g}_1^{(max)}$ and $\hat{g}_2^{(max)}$ are the sample means of the two teacher scores, $g_1$ and $g_2$, across the entire training set. Additionally, the values $\hat{g}_1^{(max)}$ and $\hat{g}_2^{(max)}$ are used to adapt the normalization parameter in (\ref{eq:student_unsupLoss}), $\lambda$, at the $i$-th iteration of the student training phase according to the following rule.

\begin{equation}
\label{eq:lambda_update}
\lambda^{(i)}=
    \begin{cases}
        \lambda^{(i-1)}+\epsilon G_2 & \text{if } G_2 \geq G_1\\
        \lambda^{(i-1)}-\epsilon G_1 & \text{otherwise,}
    \end{cases}
\end{equation} where $\epsilon$ is a fixed step size, and $G_1 \triangleq \mathbb{E}\left[\frac{\hat{g_1}^{(max)}-g_1}{\hat{g_1}^{(max)}}\right]$ and $G_2 \triangleq \mathbb{E}\left[\frac{\hat{g_2}^{(max)}-g_2}{\hat{g_2}^{(max)}}\right]$ are the average normalized reference gaps for SSNR and SINR, respectively. We, again, approximate the expectation operators by taking the sample means of the reference gaps, $\frac{\hat{g_1}^{(max)}-g_1}{\hat{g_1}^{(max)}}$ and $\frac{\hat{g_2}^{(max)}-g_2}{\hat{g_2}^{(max)}}$, across the mini-batch data points. The update equation (\ref{eq:lambda_update}) is applied as long as $\lambda^{(i)} \in [0,1]$. If $\lambda^{(i)}$ is out of the range, it is forcibly set to the closest value in the range (i.e., 0 if $\lambda^{(i)}<0$ or 1 if $\lambda^{(i)}>1$). As the teacher and student model architectures are identical, $\hat{g}_1^{(max)}$ and $\hat{g}_2^{(max)}$ represent the highest SSNR and the highest minimum SINR values that the student model can ever achieve. Therefore, from (\ref{eq:student_unsupLoss}), $L_s\in[-1,0]$. The complete teacher-student training diagram is shown in Figure \ref{fig:teach_stu}.

\section{Test Network Architectures}
\label{sec:archs}

Three CNN architectures are used to test the performance of the proposed training scheme as described below.

\begin{figure}[t]
     \centering
     \includegraphics[width=0.7\linewidth]{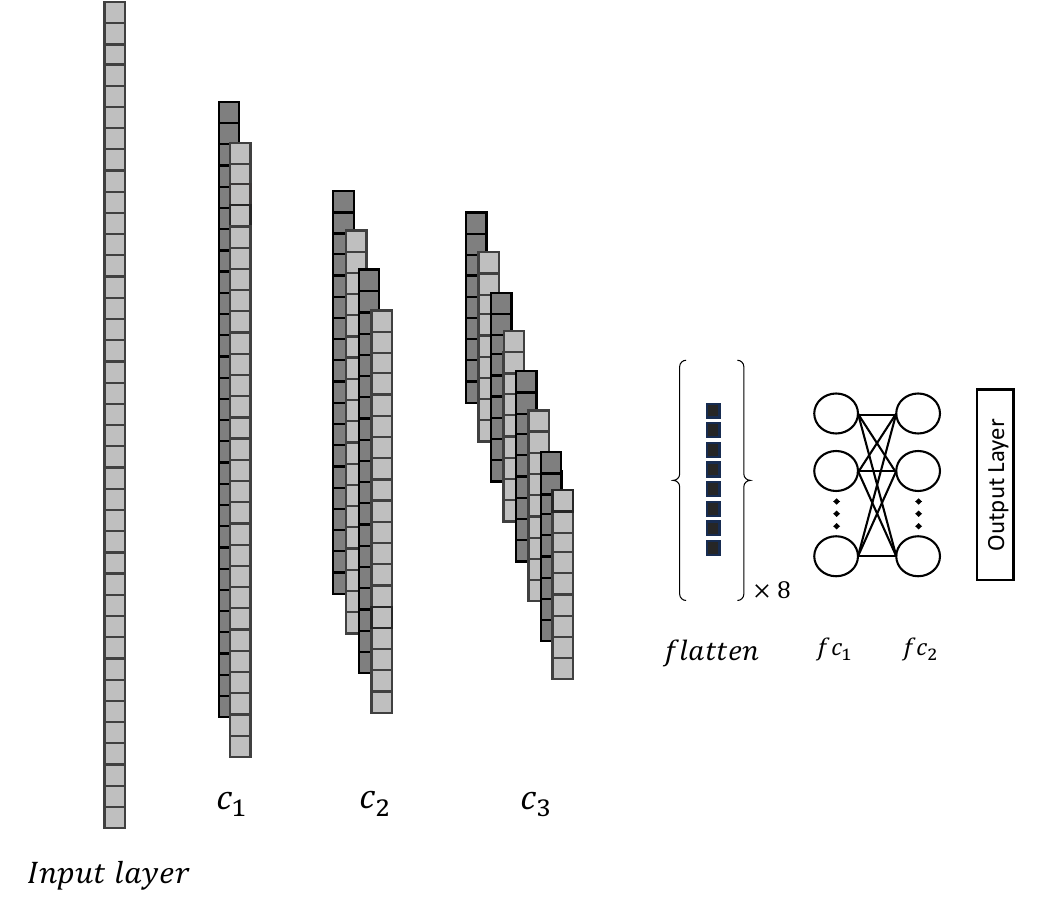}
     \caption{A 3-layer 1D-CNN example with a filter length of 11.}
     \label{fig:1DCNN}
\end{figure}

\begin{figure}[t]
     \centering
     \includegraphics[width=\linewidth]{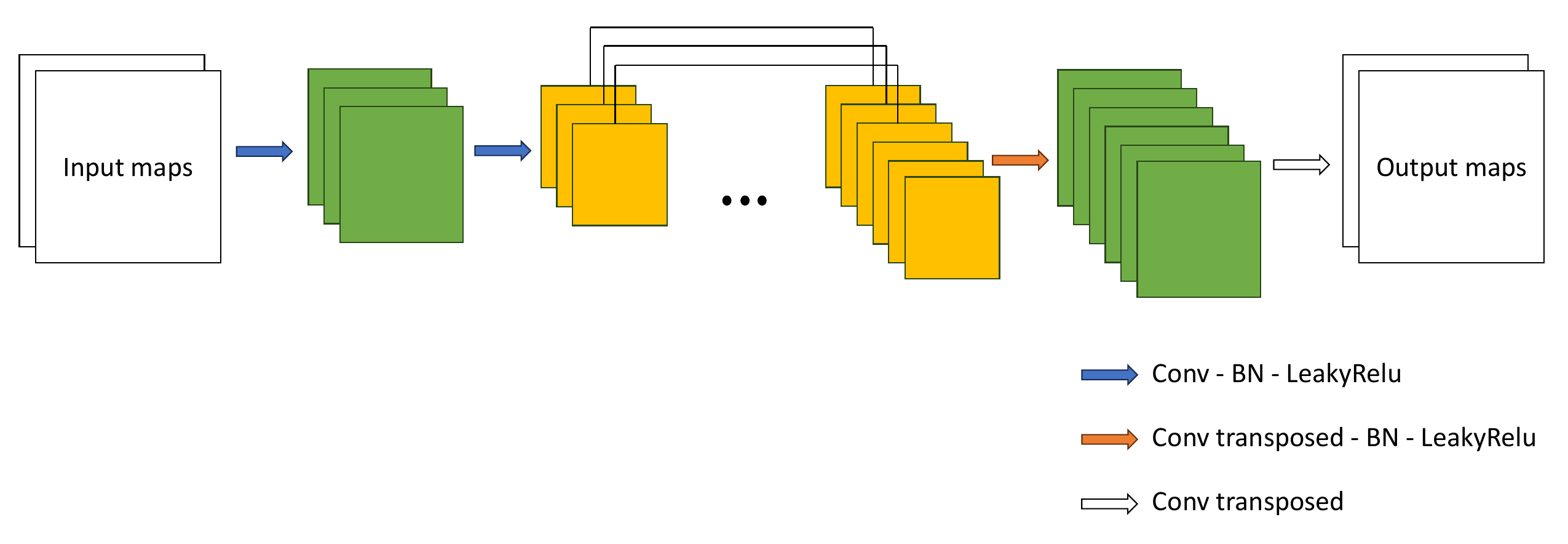}
     \caption{A convolutional autoencoder with skip-connections.}
     \label{fig:AEskipConnect}
\end{figure}

\begin{enumerate}
    \item One-dimensional CNN (1D-CNN): The first architecture is composed of three 1D convolutional layers, $c_1, c_2$ and $c_3$, and two fully connected (FC) layers, $fc_1$ and $fc_2$, as shown in Figure \ref{fig:1DCNN}. Every convolutional layer is followed by a BN layer and a LeakyReLU activation layer. Every FC layer is followed by a LeakyReLU activation layer.

    \item Convolutional Autoencoder (CAE): The second architecture is a convolutional autoencoder (CAE) with skip connections. The general layout of the CAE is shown in Figure \ref{fig:AEskipConnect}. In our experiments, the encoder and the decoder parts consist of four convolutional layers each. The skip connections are established after the second convolutional layer for all convolutional layers of the encoder. They are established by concatenating the encoder feature maps of the encoder to their respective feature maps of the decoder.

    \item U-net: The third model is an off-the-shelf model known as U-net. The detailed architecture of the original U-net is found in \cite{ronneberger2015u}. U-net was originally used for medical image segmentation. Images are usually large in size and consist of non-negative values. However, negative values are required for the output beamforming vectors and the input size is limited by the number of users and the number of antennas. Consequently, a number of modifications are made to the original U-net architecture. These modifications include dropping the pooling layers, using LeakyReLU activation function instead of ReLU, and applying zero padding to preserve the size of the input. It is worth noticing that U-net utilizes skip-connections just like the CAE.
\end{enumerate}

The 1D-CNN model input is given by

\begin{equation}
\label{eq:inputLayer1D}
     \mathcal{I}_l^{(\text{1D})} = [\mathfrak{R}(\mathbf{h}_{l1}^{\top}),...,\mathfrak{R}(\mathbf{h}_{lN}^{\top}),\mathfrak{R}(\mathbf{a}(\theta_l)^{\top}), \mathfrak{I}(\mathbf{h}_{l1}^{\top}),...,\mathfrak{I}(\mathbf{h}_{lN}^{\top}),\mathfrak{I}(\mathbf{a}(\theta_l)^{\top})]^{\top} ~ \forall l.   
\end{equation}

On the other hand, the inputs to both the CAE and the U-net models have 2D form and are identical. The 2D input is given by

\begin{equation}
\label{eq:inputLayerAE}
     \mathcal{I}_l^{(\text{CAE})} = \operatorname{blk}\left[\mathfrak{R}(\mathbf{\mathcal{H}}_l), \mathfrak{I}(\mathbf{\mathcal{H}}_l)\right] ~ \forall l,    
\end{equation}
where $\mathbf{\mathcal{H}}_l\triangleq[\mathbf{h}_{l1},...,\mathbf{h}_{lN},\mathbf{a}(\theta_l)]$ is the CSI of the $l$-th AP. 

The direct outputs of the $l$-th network are constructed similarly to the input forms of the 1D-CNN and the CAE (U-net) cases. In other words, the output for the 1D-CNN is denoted by

\begin{equation}
\label{eq:outputLayer1D}
     \boldsymbol{\omega}_l = [\boldsymbol{\omega}_l^{(r)\top}, \boldsymbol{\omega}_l^{(i)\top}]^{\top} ~ \forall l, 
\end{equation}
and the output of the CAE and U-net is constructed as
\begin{equation}
\label{eq:outputLayerAE}
     \boldsymbol{\Omega}_l = \operatorname{blk}\left[\boldsymbol{\Omega}_l^{(r)}, \boldsymbol{\Omega}_l^{(i)}\right] ~ \forall l,    
\end{equation} where the superscripts $(r)$ and $(i)$ indicate the real and imaginary parts of the complex prediction, respectively.


To enforce the power constraints, the final complex predictions of the 1D-CNN and the CAE (U-net) are given by

\begin{equation}
\label{eq:normalization1D}
          \Bar{\boldsymbol{\omega}}_l = \sqrt{P_l} \frac{\boldsymbol{\omega}_l^{(r)}+j\boldsymbol{\omega}_l^{(i)}}{||\boldsymbol{\omega}_l^{(r)}+j\boldsymbol{\omega}_l^{(i)}||} ~\forall l
\end{equation}
and
\begin{equation}
\label{eq:normalizationAE}
          \Bar{\boldsymbol{\Omega}}_l = \sqrt{P_l} \frac{\boldsymbol{\Omega}_l^{(r)}+j\boldsymbol{\Omega}_l^{(i)}}{||\boldsymbol{\Omega}_l^{(r)}+j\boldsymbol{\Omega}_l^{(i)}||} ~\forall l,
\end{equation} respectively.

The post-processed outputs are constructed as $\Bar{\boldsymbol{\omega}}_l\triangleq[\mathbf{w}_{l1}^{\top},...,\mathbf{w}_{l(N+1)}^{\top}]^{\top}$ and $\Bar{\boldsymbol{\Omega}}_l\triangleq[\mathbf{w}_{l1},...,\mathbf{w}_{l(N+1)}]^{\top}$. As this is an unsupervised approach, a ground truth for $\Bar{\boldsymbol{\omega}}_l$ or $\Bar{\boldsymbol{\Omega}}_l$ is not present. Instead, the SSNR and SINR are calculated to evaluate the loss, which was discussed in Section \ref{sec:loss}.




\section{Dataset Generation}
\label{sec:dataset}

The system parameters, at which the unsupervised datasets are generated, follow in general the system parameters used to evaluate the joint beamforming performance in \cite{demirhan2023cell}. That is, we consider 2 APs, 16 antennas per AP, a power budget of 1W per AP, a noise power of 1W for both the UEs and the receiver, and a sensing channel gain power of 0.1. Depending on the orientation of the target with respect to the APs, the sensing channel is generated according to (\ref{eq:steering}). Equation (\ref{eq:steering}) is also used to generate the communication channels given the orientation of UEs with respect to the APs. Similar ISAC channel models were considered in \cite{demirhan2023cell,qi2022hybrid,zhuo2023performance}. Only line of sight is considered for the channel model.

    



\begin{figure}[t]
    \centering
    
    \subfloat[\label{subfig:pos1}Pos-1]{\includegraphics[width=0.5\linewidth]{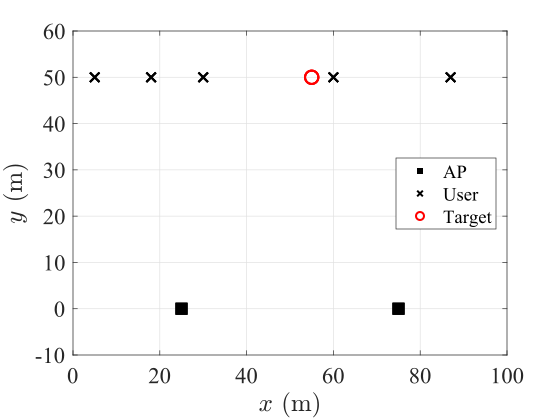}}
    \subfloat[\label{subfig:pos2}Pos-2]{\includegraphics[width=0.5\linewidth]{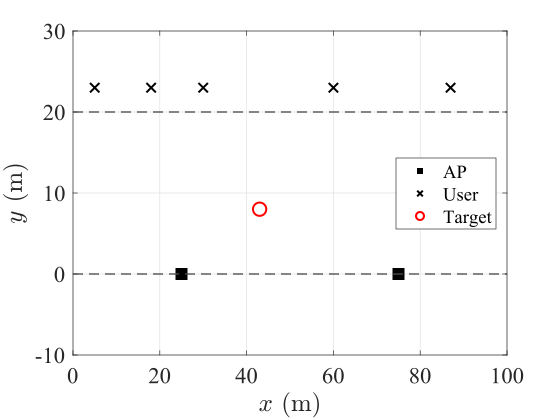}}

    \caption{Different position distribution schemes used for dataset generation at $L=2$ and $N=5$. The area enclosed by the dashed line is the road, which indicates the possible locations of the ST.}
    \label{fig:pos}
\end{figure}

Figure \ref{fig:pos} shows two position distribution configurations for the UEs and the target. Specifically, Figure \ref{subfig:pos1} shows an example of the distribution scheme followed by \cite{demirhan2023cell} at $L=2$ and $N=5$, where the $y$-coordinate for all agents is fixed at 50m whereas the $x$-coordinates for the agents are randomized between $x=0$ and $x=100$m. The scheme in Figure \ref{subfig:pos2} is more closely related to the realistic scenario depicted in Figure \ref{fig:isacSetup}, where a two-lane road with a total width of 20m is considered. While the UE positions are fixed at $y=23$m (3m away from the traffic) and randomized along the $x$-axis between 0 and 100m, both $x$ and $y$ coordinates are randomized for the target between $x=0$ and $x=100$m, and $y=0$ and $y=20$m. In both configurations, the 2 APs are positioned at (25,0)m and (75,0)m. To avoid confusion, we refer to the position distribution configuration of the benchmark (i.e., Figure \ref{subfig:pos1}) by Pos-1. Our position distribution configuration (i.e., Figure \ref{subfig:pos2}) is referred to by Pos-2. Furthermore, we specify at each experiment which position distribution configuration is used for the corresponding dataset generation. All random positions are drawn from uniform distributions.

After generating the agent positions, the communication and sensing channels are both calculated according to (\ref{eq:steering}) as mentioned earlier. Once the channels are generated, the unsupervised dataset generation step is accomplished. On the other hand, following a supervised training scheme requires evaluating the beamforming vectors of all generated data points, which consumes a tremendous amount of time to generate a sufficient number of data points for training.


Unless mentioned otherwise, 20,000 point-datasets are generated, where 97\% is used for training and the remaining points are used to validate the model's performance after every training epoch.


\section{Results and Discussion}
\label{sec:results}

This section investigates the performance of the proposed DL unsupervised scheme. First, the performance of the algorithm at different networks is investigated by observing how training and validation SINR and SSNR evolve throughout the training process, and comparing the resulting curves to the mean performance of the CVX-based solution \cite{demirhan2023cell}. Next, a sensitivity assessment is conducted to study the effect of the parameters $\beta$ and $\lambda$ on the training performance. Then, the effects of changing the number of UEs on the algorithm's performance is studied. We conclude the experiments by a comparison between the proposed DL algorithm and the CVX-based solution of \cite{demirhan2023cell} in terms of running time.

We use a single 2GB GPU NVIDIA Quadro P600 to train the models. The run time experiment is conducted using an 11th generation Intel CPU with an i5 core running at 2.40 GHz. The Pytorch library is used to conduct all experiments. Both SINR teacher and student models are trained for a maximum of 1000 epochs, and the training is terminated if no improvement in the validation metrics occurs for 100 epochs (i.e., patience is set to 100). SSNR teachers are trained for 100 epochs without early stopping. For student training, the initial value of the normalization parameter, $\lambda^{(0)}$, is set to 0.5. Also, the step size, $\epsilon$, in (\ref{eq:lambda_update}) is set to 0.01. A mini-batch size of 500 is used for all training processes. ADAM optimizer is used for training all models at an initial learning rate of 0.01. The learning rate is reduced by a factor of 10 if no improvement in the average unsupervised loss through the epoch is observed for 10 consecutive epochs (i.e., ReduceLROnPlateau learning rate scheduling module in Pytorch library). The specifications for the three models used to assist the algorithm at $N=5$ UEs and $M=16$ antennas per AP are are detailed in Table \ref{table:CNNparams}. 



\begin{table}[t]
\centering
\caption{Parameter summary for the three test CNNs.}
\label{table:CNNparams}
\resizebox{\linewidth}{!}{%
\begin{tabular}{|>{\hspace{0pt}}m{0.188\linewidth}|>{\centering\hspace{0pt}}m{0.296\linewidth}|ll>{\centering\arraybackslash\hspace{0pt}}m{0.244\linewidth}|} 
\hline
\multicolumn{5}{|>{\centering\arraybackslash\hspace{0pt}}m{0.925\linewidth}|}{1D-CNN}                                                                                                                                                                                    \\ 
\hline
Conv
  layers                                                    & Channels & \multicolumn{2}{>{\centering\hspace{0pt}}m{0.198\linewidth}|}{Filter
  size} & Padding                                                                                                     \\ 
\hline
\multicolumn{1}{|>{\centering\hspace{0pt}}m{0.188\linewidth}|}{$c_1$} & 2        & \multicolumn{2}{>{\centering\hspace{0pt}}m{0.198\linewidth}|}{(11,1)}        & (0,0)                                                                                                       \\ 
\hline
\multicolumn{1}{|>{\centering\hspace{0pt}}m{0.188\linewidth}|}{$c_2$} & 4        & \multicolumn{2}{>{\centering\hspace{0pt}}m{0.198\linewidth}|}{(11,1)}        & (0,0)                                                                                                       \\ 
\hline
\multicolumn{1}{|>{\centering\hspace{0pt}}m{0.188\linewidth}|}{$c_3$} & 8        & \multicolumn{2}{>{\centering\hspace{0pt}}m{0.198\linewidth}|}{(11,1)}        & (0,0)                                                                                                       \\ 
\hline
FC
  layers                                                      & \multicolumn{4}{>{\centering\arraybackslash\hspace{0pt}}m{0.738\linewidth}|}{Number
  of neurons}                                                                                                     \\ 
\hline
\multicolumn{1}{|>{\centering\hspace{0pt}}m{0.188\linewidth}|}{$fc1$} & \multicolumn{4}{>{\centering\arraybackslash\hspace{0pt}}m{0.738\linewidth}|}{90}                                                                                                                      \\ 
\hline
\multicolumn{1}{|>{\centering\hspace{0pt}}m{0.188\linewidth}|}{$fc2$} & \multicolumn{4}{>{\centering\arraybackslash\hspace{0pt}}m{0.738\linewidth}|}{90}                                                                                                                      \\ 
\hline
\multicolumn{5}{|>{\centering\arraybackslash\hspace{0pt}}m{0.925\linewidth}|}{CAE}                                                                                                                                                                                       \\ 
\hline
Parameter                                                        & \multicolumn{2}{>{\centering\hspace{0pt}}m{0.397\linewidth}|}{Encoder}                    & \multicolumn{2}{>{\centering\arraybackslash\hspace{0pt}}m{0.339\linewidth}|}{Decoder}                     \\ 
\hline
Channels                                                         & \multicolumn{2}{>{\centering\hspace{0pt}}m{0.397\linewidth}|}{16, 32, 64 and 128}         & \multicolumn{2}{>{\centering\arraybackslash\hspace{0pt}}m{0.339\linewidth}|}{32, 64,128 and 256}          \\ 
\hline
Filter size                                                      & \multicolumn{2}{>{\centering\hspace{0pt}}m{0.397\linewidth}|}{(3,5), (2,5), (1,3), (1,3)} & \multicolumn{2}{>{\centering\arraybackslash\hspace{0pt}}m{0.339\linewidth}|}{(3,5), (2,5), (1,3), (1,3)}  \\ 
\hline
Padding                                                          & \multicolumn{2}{>{\centering\hspace{0pt}}m{0.397\linewidth}|}{0}                          & \multicolumn{2}{>{\centering\arraybackslash\hspace{0pt}}m{0.339\linewidth}|}{0}                           \\ 
\hline
\multicolumn{5}{|>{\centering\arraybackslash\hspace{0pt}}m{0.925\linewidth}|}{U-net}                                                                                                                                                                                     \\ 
\hline
Parameter                                                        & \multicolumn{2}{>{\centering\hspace{0pt}}m{0.397\linewidth}|}{Encoder}                    & \multicolumn{2}{>{\centering\arraybackslash\hspace{0pt}}m{0.339\linewidth}|}{Decoder}                     \\ 
\hline
Channels                                                         & \multicolumn{2}{>{\centering\hspace{0pt}}m{0.397\linewidth}|}{16, 32, 64 and 128}         & \multicolumn{2}{>{\centering\arraybackslash\hspace{0pt}}m{0.339\linewidth}|}{32, 64,128 and 256}          \\ 
\hline
Filter size                                                      & \multicolumn{2}{>{\centering\hspace{0pt}}m{0.397\linewidth}|}{(3,3)}                      & \multicolumn{2}{>{\centering\arraybackslash\hspace{0pt}}m{0.339\linewidth}|}{(3,3)}                       \\ 
\hline
Padding                                                          & \multicolumn{2}{>{\centering\hspace{0pt}}m{0.397\linewidth}|}{(1,1)}                      & \multicolumn{2}{>{\centering\arraybackslash\hspace{0pt}}m{0.339\linewidth}|}{(1,1)}                       \\
\hline
\end{tabular}
}
\end{table}

The benchmark performance is obtained by solving problem (\ref{eq:optPriority}) at $\rho=0.5$, where the sensing beam is designed by projecting its vector to the null space of the communication channel. The obtained $\gamma_{high}$ from the bisection algorithm is used for solving problem (\ref{eq:jsc2}). The problem is solved for 200 points randomly selected from the corresponding dataset, whereas the average SSNR and SINR of these points are the benchmark for our method.

\subsection{Benchmark comparison with teacher-student scheme}

\begin{figure}[t]
    \centering
    
    \subfloat[\label{subfig:BenchSSNR1D}1D-CNN]{\includegraphics[width=0.33\linewidth]{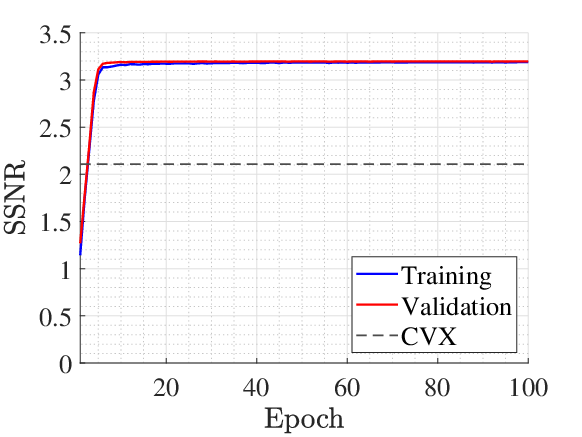}}
    \subfloat[\label{subfig:BenchSSNRAE}CAE]{\includegraphics[width=0.33\linewidth]{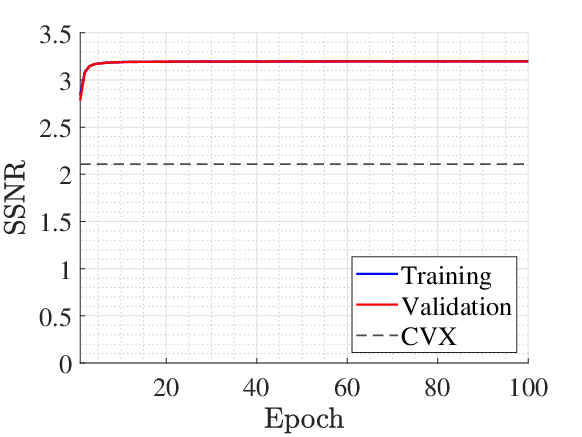}}
    \subfloat[\label{subfig:BenchSSNRUNet}U-net]{\includegraphics[width=0.33\linewidth]{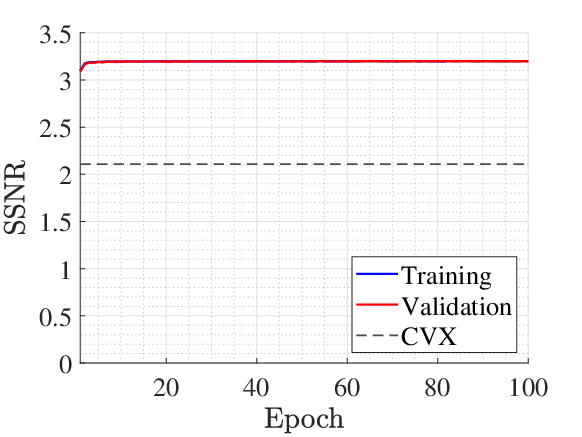}}

    \caption{SSNR teacher performance.}
    \label{fig:BenchSSNRTeach}
    
\end{figure}

All results in this section are based on datasets generated via Pos-1 configuration (Refer to Figure \ref{subfig:pos1}). Figure \ref{fig:BenchSSNRTeach} shows SSNR teacher performance through training and validation using the three networks. It can be seen from both Figure \ref{fig:BenchSSNRTeach} and equation (\ref{eq:SSNR}) that maximizing the SSNR alone is relatively simple. All networks attain the same level of SSNR within 100 epochs of training. This level surpasses the CVX-based solution (i.e., the dashed line in Figure \ref{fig:BenchSSNRTeach}) by a considerable margin. The only difference between the three networks in terms of the SSNR curves is the speed at which the maximum SSNR is attained, where U-net is the fastest to converge and the 1D-CNN is the slowest. Notice that the initial points of the three training curves in Figure \ref{fig:BenchSSNRTeach} are different, since the first training point is always recorded at the end of the first epoch.

\begin{figure}[t]
    \centering
    
    \subfloat[\label{subfig:BenchSINR1D}1D-CNN]{\includegraphics[width=0.33\linewidth]{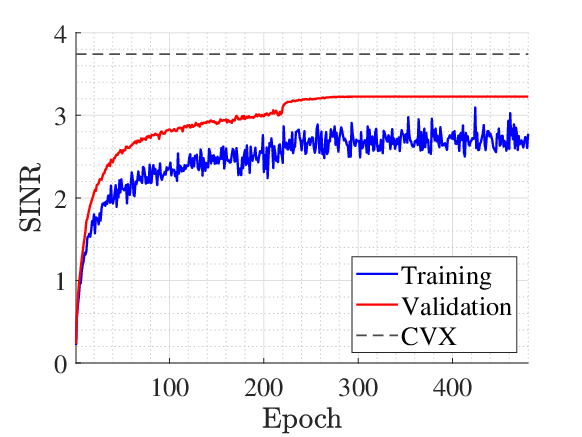}}
    \subfloat[\label{subfig:BenchSINRAE}CAE]{\includegraphics[width=0.33\linewidth]{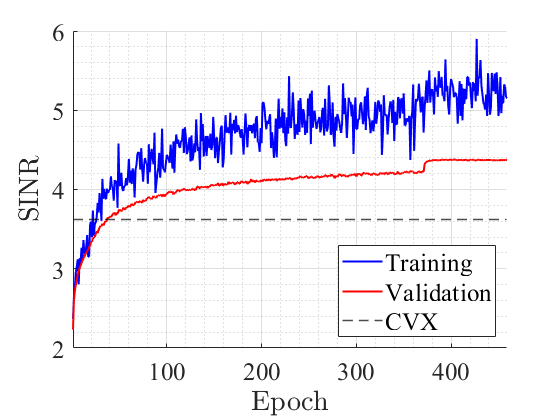}}
    \subfloat[\label{subfig:BenchSINRUNet}U-net]{\includegraphics[width=0.33\linewidth]{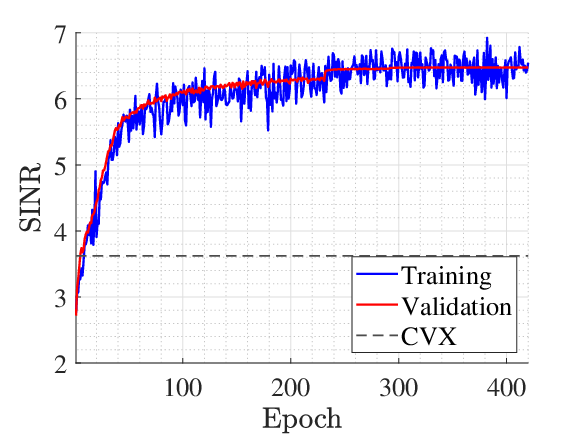}}

    \caption{SINR teacher performance.}
    \label{fig:BenchSINRTeach}
\end{figure}

Unlike the concave quadratic nature of the objective that SSNR teachers are expected to optimize, maximizing the minimum SINR is a more difficult task due to its non-convex nature. Figure \ref{fig:BenchSINRTeach} shows clear distinctions between the three test networks in terms of effectiveness in maximizing the minimum SINR. While both CAE and U-net could achieve higher SINR than the CVX-based algorithm (represented by the dashed lines in Figure \ref{fig:BenchSINRTeach}) in terms of both training and validation, 1D-CNN validation curve barely attains the CVX-based algorithm performance as shown in Figure \ref{subfig:BenchSINR1D}. Consequently, the student's performance of the 1D-CNN will not be able to jointly maximize SSNR and SINR properly, since its maximum performance is bounded by the teachers' performance.

\begin{figure}[t]
    \centering
    
    \subfloat[\label{subfig:BenchLossStuAE}]{\includegraphics[width=0.5\linewidth]{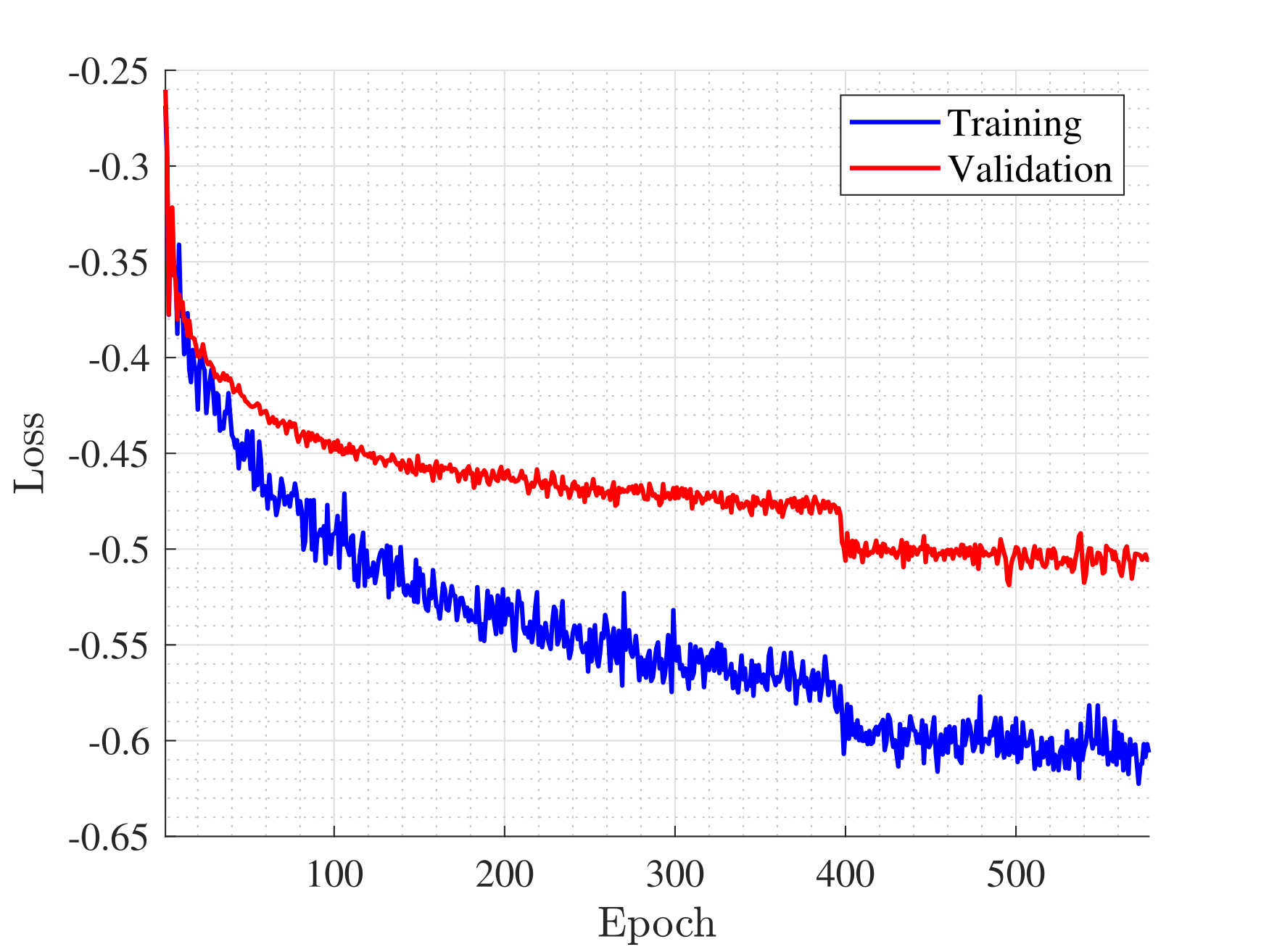}}
    \subfloat[\label{subfig:BenchLamStuAE}]{\includegraphics[width=0.5\linewidth]{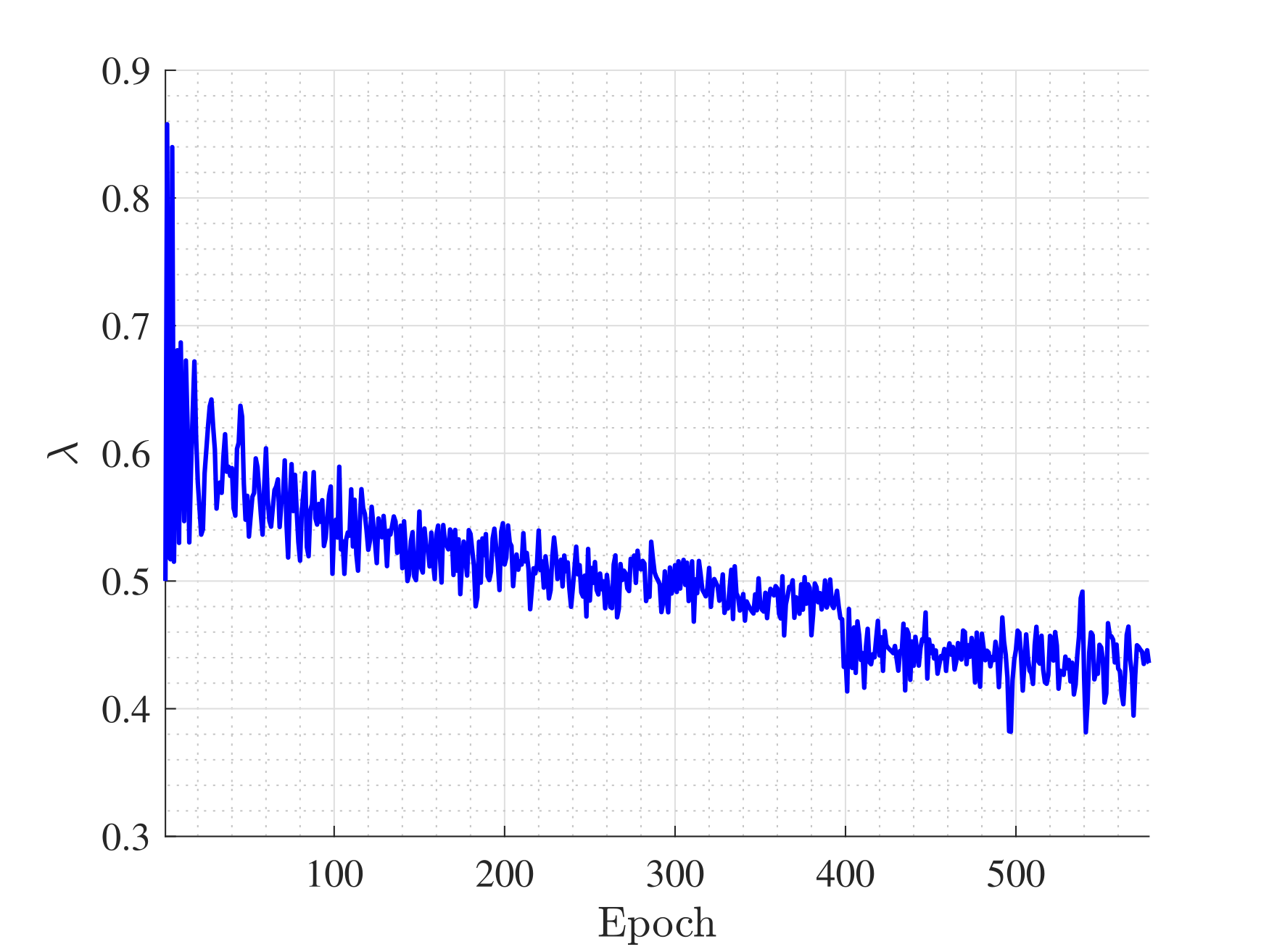}}
    
    \subfloat[\label{subfig:BenchSINRStuAE}]{\includegraphics[width=0.5\linewidth]{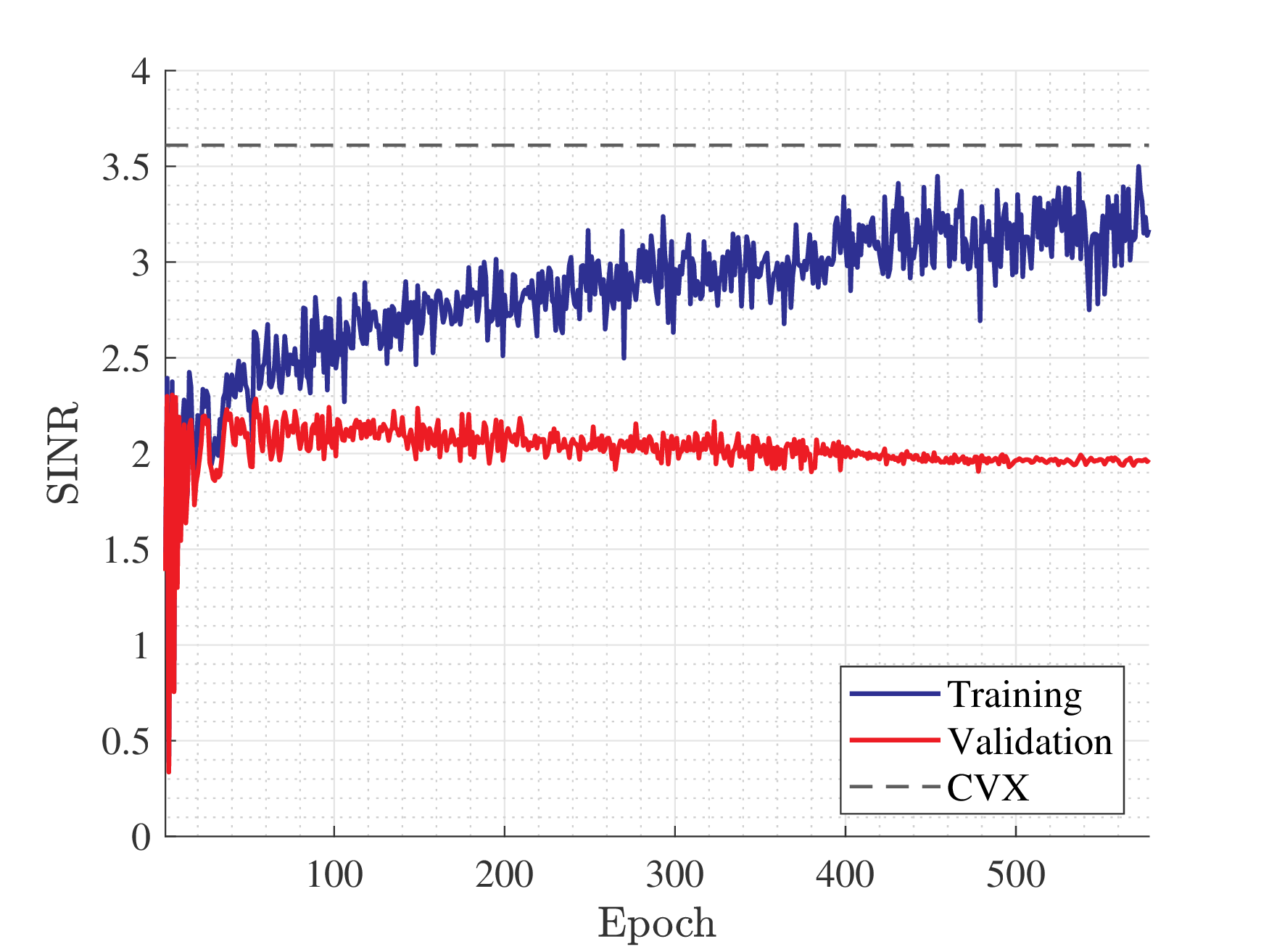}}
    \subfloat[\label{subfig:BenchSSNRStuAE}]{\includegraphics[width=0.5\linewidth]{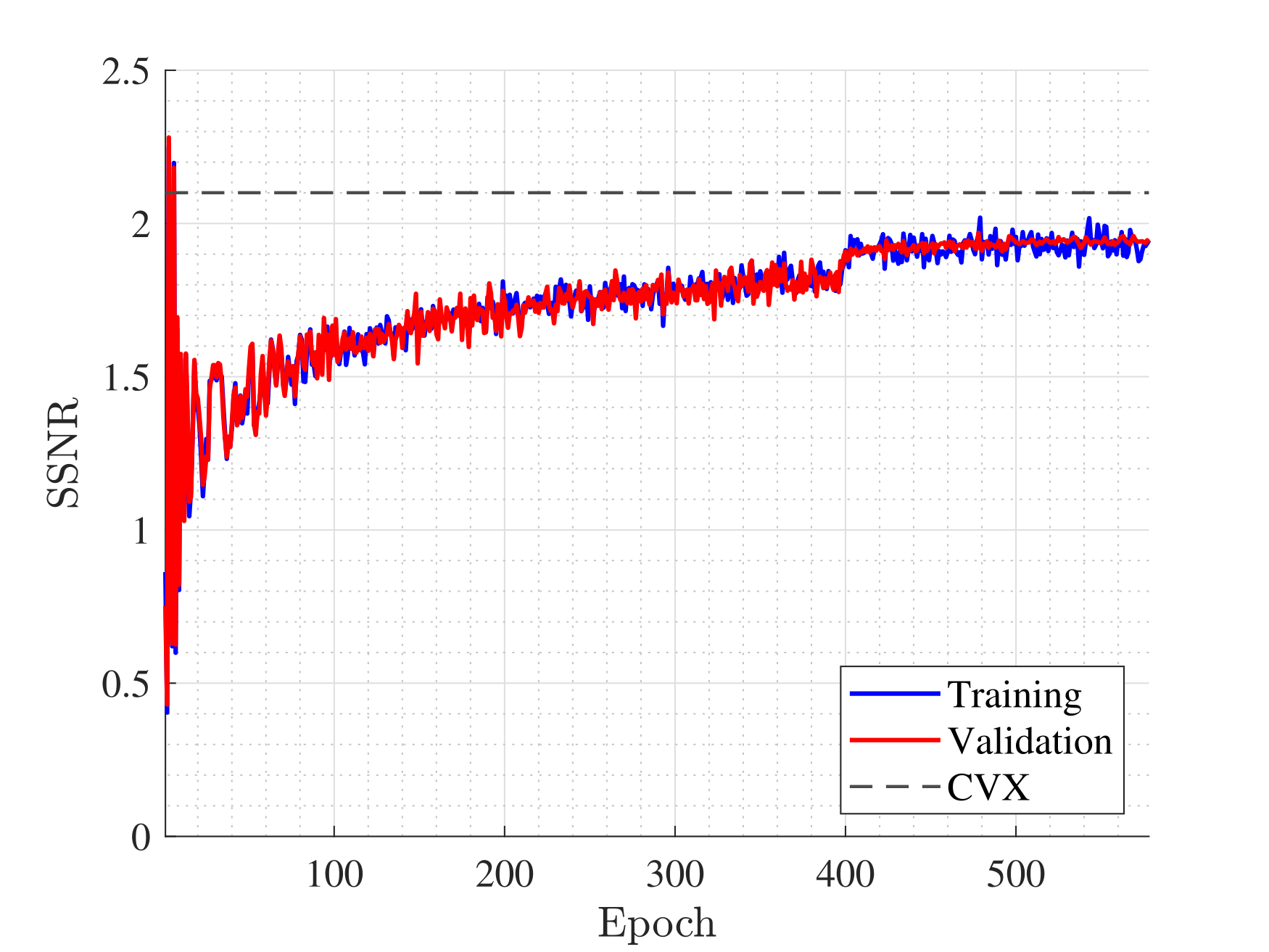}}

    \caption{Student performance of the CAE.}
    \label{fig:BenchStudentAE}
\end{figure}

The training-validation curves of the CAE in Figure \ref{subfig:BenchSINRAE} exhibit an overfitting issue, which reflects poorly on the student's performance depicted in Figure \ref{subfig:BenchSINRStuAE}. U-net does not suffer from any of the issues exhibited by the CAE and the 1D-CNN. 

\begin{figure}[t]
    \centering
    
    \subfloat[\label{subfig:BenchLossStuUnet}]{\includegraphics[width=0.5\linewidth]{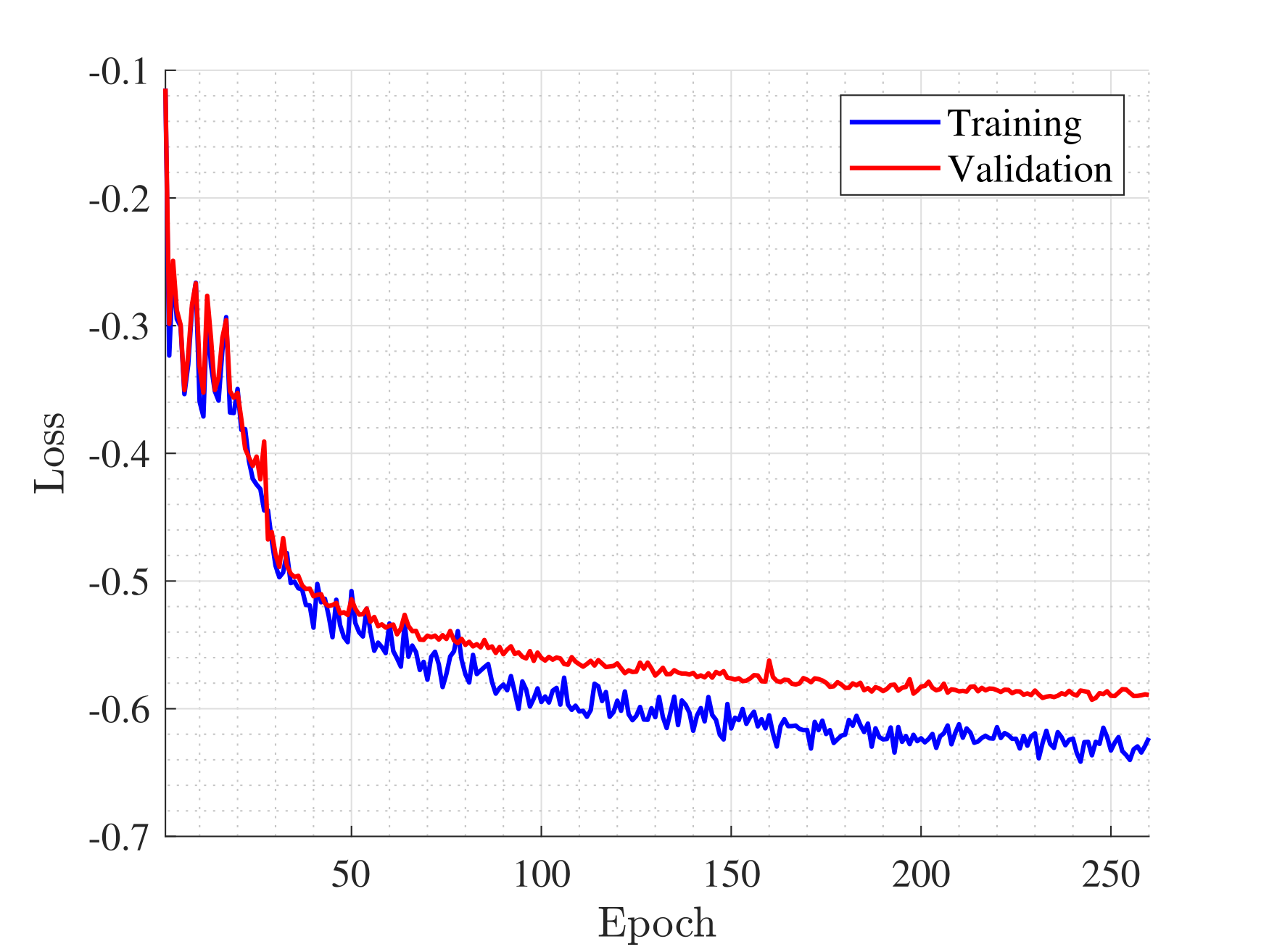}}
    \subfloat[\label{subfig:BenchLamStuUnet}]{\includegraphics[width=0.5\linewidth]{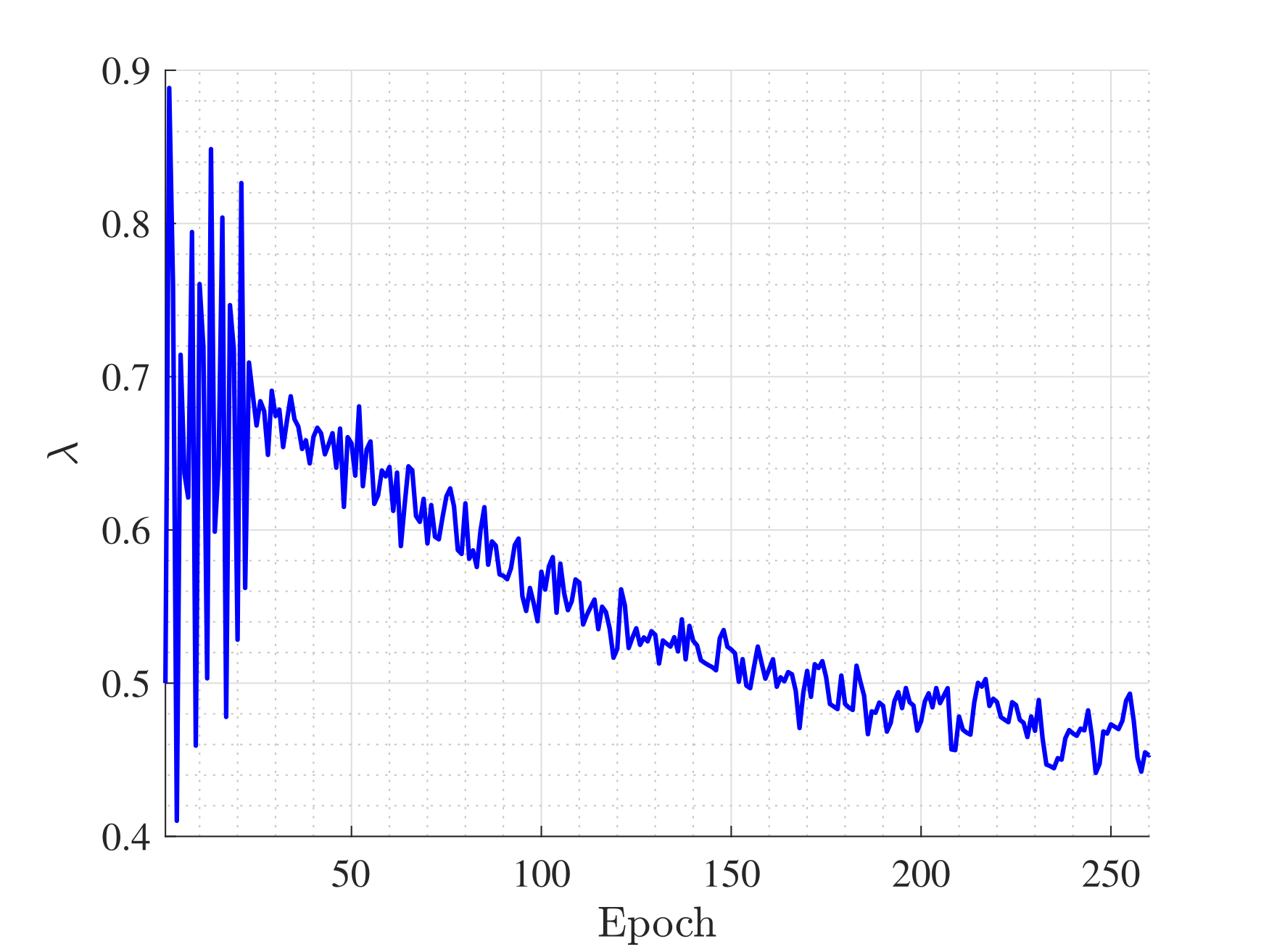}}
    
    \subfloat[\label{subfig:BenchSINRStuUnet}]{\includegraphics[width=0.5\linewidth]{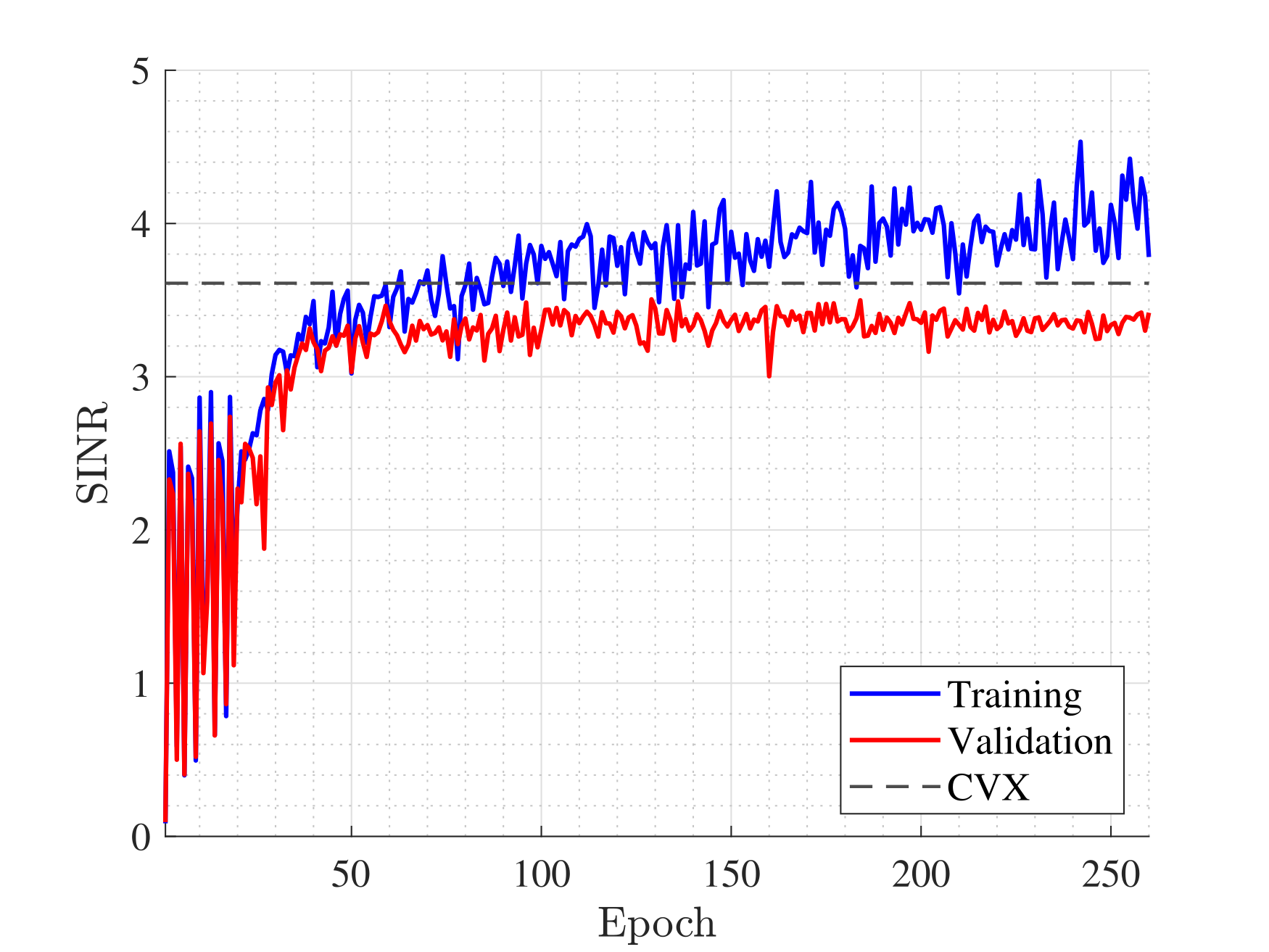}}
    \subfloat[\label{subfig:BenchSSNRStuUnet}]{\includegraphics[width=0.5\linewidth]{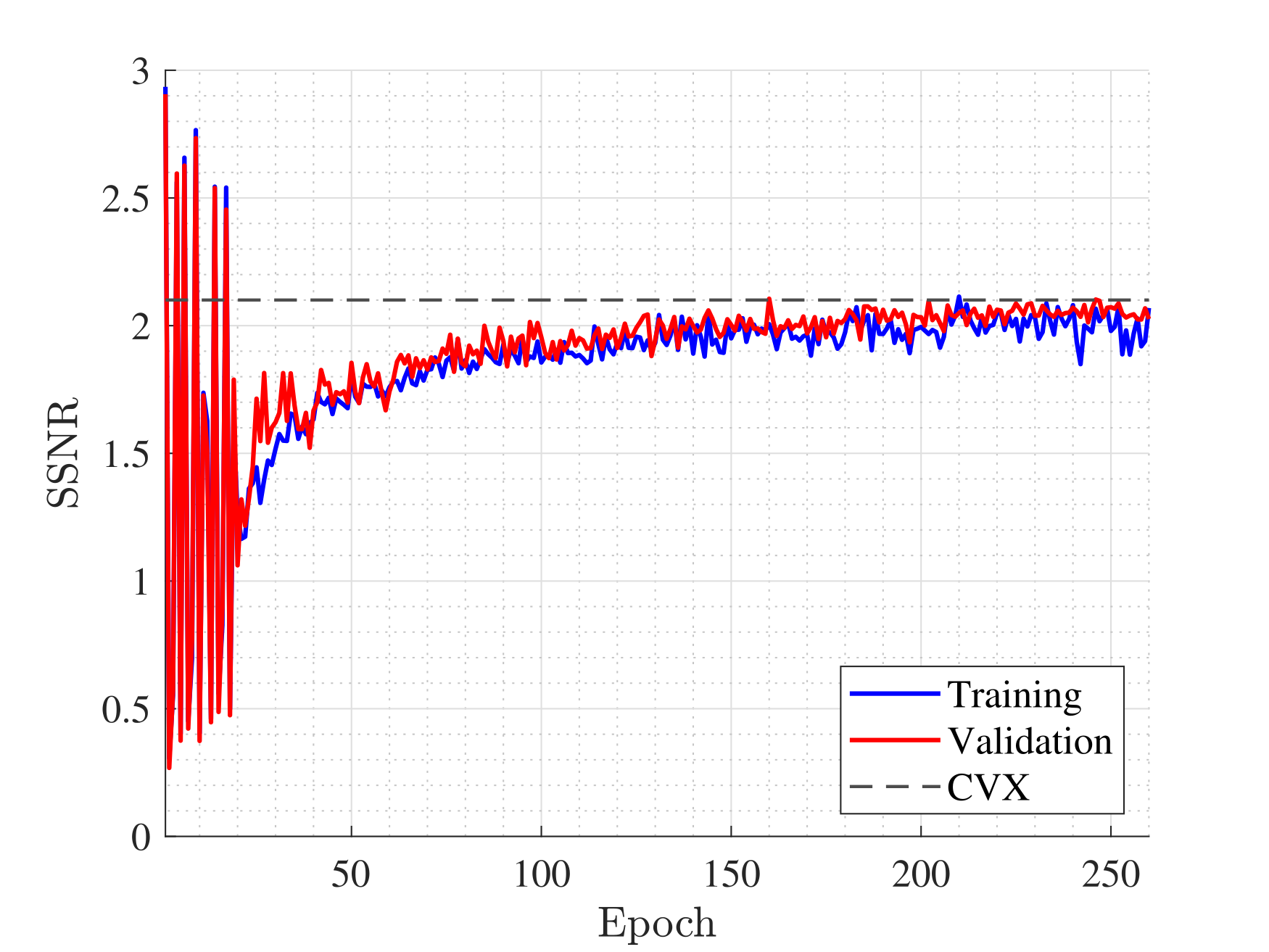}}

    \caption{Student performance of U-net.}
    \label{fig:BenchStudentUNet}

\end{figure}

Figure \ref{subfig:BenchSSNRUNet} and Figure \ref{subfig:BenchSINRUNet} show that the training-validation curves of the U-net teacher models converge to considerably higher SSNR and SINR than the CVX-based performance. Thus, it is expected that the U-net student can achieve promising results. Said results are shown in Figure \ref{subfig:BenchSINRStuUnet} and Figure \ref{subfig:BenchSSNRStuUnet}. It is expected that the student's final performance should be at least slightly lower than the CVX-based solution, since the CVX-based solution is near-optimal whereas DNNs are general approximators. The trained student models, however, can evaluate the beamforming vectors at every AP in real time given CSI. However, the CVX-based solution burdens the CPU with calculating the beamforming vectors using a time consuming iterative algorithm, and sending the evaluated beamforming vectors to their respective APs. On the other hand, the trained student models only require the corresponding AP's CSI to evaluate beamforming vectors that are close in quality to the CVX-based solution, and can be deployed in real-time scenarios.

\subsection{Sensitivity assessment}

\begin{figure}[t]
     \centering
     \includegraphics[width=0.6\linewidth]{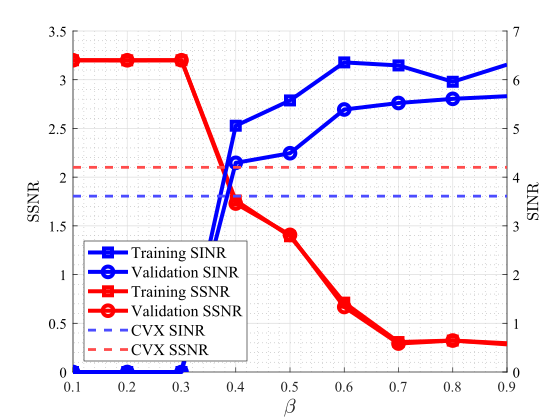}
     \caption{SINR and SSNR at fixed $\beta$.}
     \label{fig:assessmentBeta}
\end{figure}

\begin{figure}[!t]
     \centering
     \includegraphics[width=0.6\linewidth]{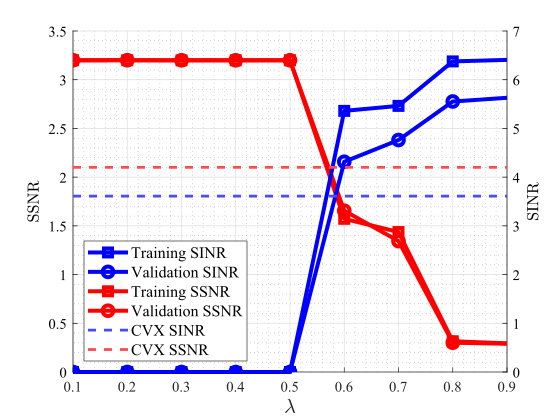}
     \caption{SINR and SSNR at fixed $\lambda$.}
     \label{fig:assessmentLambda}
\end{figure}

All results in this section are based on datasets generated via Pos-1 configuration (Refer to Figure \ref{subfig:pos1}). One possible solution to the joint optimization problem using unsupervised learning is to directly select a fixed $\beta$ in (\ref{eq:teacher_unsupLoss}) using a grid search. Figure \ref{fig:assessmentBeta} shows the expected behavior of U-net models trained using $L_1$ directly at a fixed $\beta$ thorough the entire training process. It can be seen from Figure \ref{fig:assessmentBeta} that only a narrow range of $\beta$ values can achieve similar performance to the teacher-student scheme. It should be also noticed that a considerably large range of $\beta$ results in a biased performance towards SINR or SSNR. This is not equivalent to the parameter $\rho$, which gradually controls the tradeoff between sensing and communication \cite{demirhan2023cell}. Additionally, it is expected that the curves in Figure \ref{fig:assessmentBeta} are susceptible to uncontrollable changes caused by changes in the system model, network architecture, training parameters, or training data points. Such susceptibility to minor changes renders the fixed $\beta$ training practise impractical. A similar argument can be stated about the impracticality of fixing $\lambda$ throughout the training process, whose behavior is depicted in Figure \ref{fig:assessmentLambda}. 



\subsection{Changing agent position distribution scheme and the number of UEs}

\begin{figure}[t]
    \centering
    
    \subfloat[\label{subfig:diffNSINR}]{\includegraphics[width=0.5\linewidth]{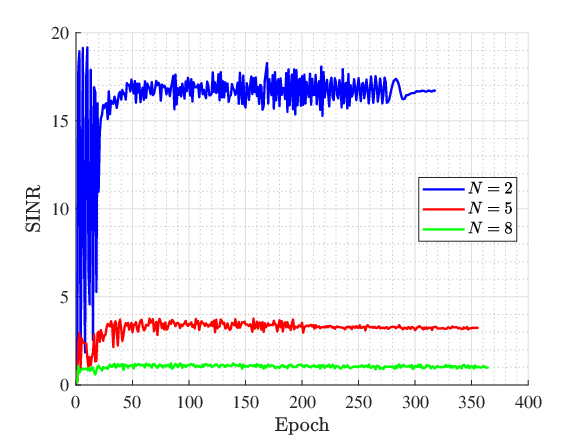}}
    \subfloat[\label{subfig:diffNSSNR}]{\includegraphics[width=0.5\linewidth]{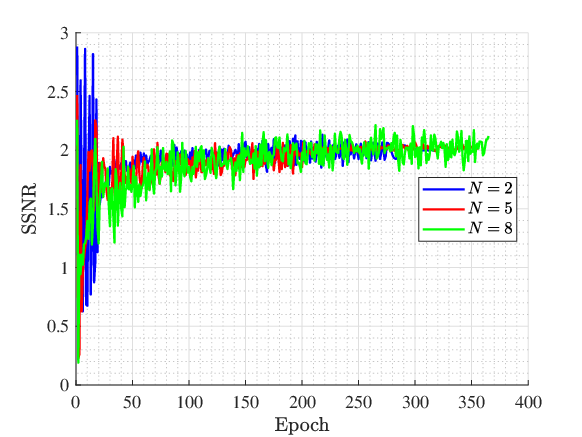}}

    \caption{Student performance of U-net at different number of UEs: (a) SINR and (b) SSNR validation curves.}
    \label{fig:diffN}

\end{figure}

We use U-net to evaluate the teacher-student scheme performance at different number of users. In this experiment, the datasets are generated using Pos-2 scheme (c.f., Figure \ref{subfig:pos2}). To this end, the validation curves for SINR and SSNR at $N=2,5,8$ are shown in Figure \ref{fig:diffN}.

It is expected that SINR would increase as $N$ decreases, since the multi-antenna equipped APs are required to serve less number of users by the same power budget. This explains the differences between the three curves in Figure \ref{subfig:diffNSINR}. On the other hand, the maximum attainable SSNR is expected to remain the same under the same power budget and sensing channel model regardless of the number of users, since it is dedicating all beams (i.e., the entire power budget) to detecting the target. This explains the similarity between the curves in Figure \ref{subfig:diffNSSNR}.

\begin{figure}[t]
    \centering
    
    \subfloat[\label{subfig:diffN5posSINR}]{\includegraphics[width=0.5\linewidth]{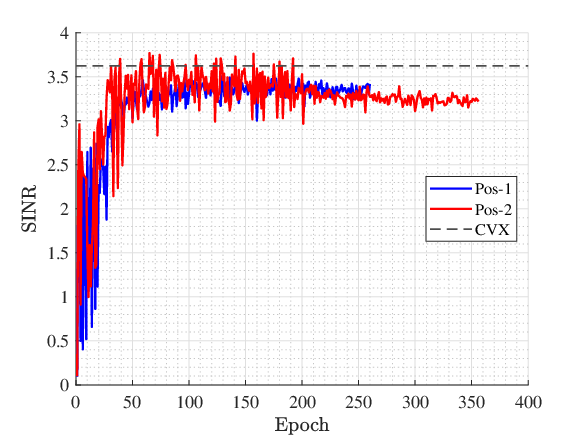}}
    \subfloat[\label{subfig:diffN5posSSNR}]{\includegraphics[width=0.5\linewidth]{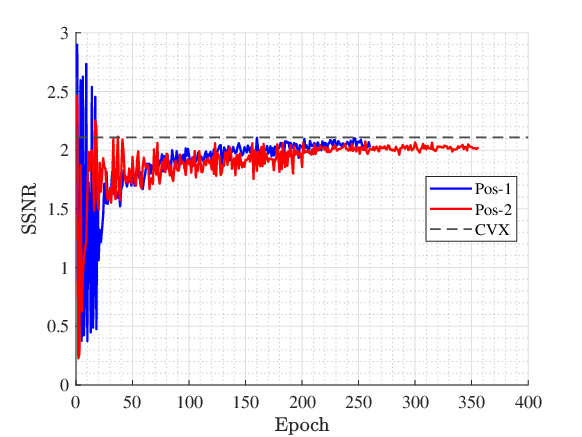}}

    \caption{Student performance at $N=5$ for different agent position distribution schemes (c.f., Figure \ref{fig:pos}): (a) SINR and (b) SSNR validation curves.}
    \label{fig:diffN5pos}

\end{figure}

\begin{figure}[!t]
    \centering
    
    \subfloat[\label{subfig:N8SINR}]{\includegraphics[width=0.5\linewidth]{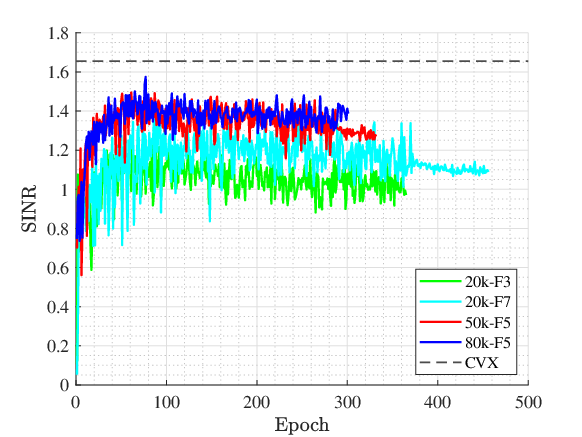}}
    \subfloat[\label{subfig:N8SSNR}]{\includegraphics[width=0.5\linewidth]{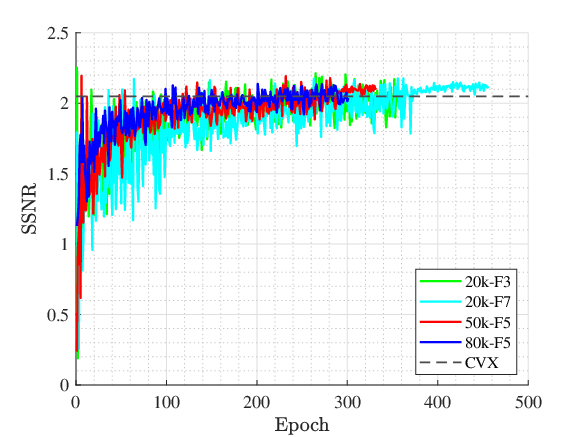}}

    \caption{Student performance at $N=8$ for different training and network setups: (a) SINR and (b) SSNR validation curves.}
    \label{fig:diffN8setups}

\end{figure}

Figure \ref{fig:diffN5pos} compares the student performance on two datasets. One dataset is generated using Pos-1 configuration and the other one is generated using Pos-2 configuration. Since the channel model depends solely on the orientation, only minor changes are expected. Said changes can be observed in Figure \ref{subfig:diffN5posSINR}, where the SINR curve declines after around 200 epochs, suggesting a slight overfitting during the training process. 

To address the results shown by Figure \ref{fig:diffN8setups}, we the notation, F$n$, to indicate a filter size of $(n, n)$ and an appropriate padding that will conserve the output size. For instance, all past U-net experiments used a filter size of $(3, 3)$ and a padding of $(1, 1)$. This filtering configuration is referred to as F3 in Figure \ref{fig:diffN8setups}. Consequently, F5 refers to a filter size of $(5, 5)$ and a padding of $(2, 2)$. Finally, F7 refers to a filter size of $(7, 7)$ and a padding of $(3, 3)$.

The observed slight overfitting in Figure \ref{subfig:diffN5posSINR} becomes more severe as the number of users increases as shown in Figure \ref{subfig:N8SINR}. This problem can be alleviated by changing the filter size of the U-net and, more importantly, by increasing the training size. The blue curve in Figure \ref{subfig:N8SINR} shows that F5 configuration with a dataset size of 80,000 points considerably enhances the SINR performance of the student model. In all cases, the overfitting issue does not affect the SSNR performance as shown in Figure \ref{subfig:diffN5posSSNR} and Figure \ref{subfig:N8SSNR}. Judging by the similar performance of the blue and red curve corresponding schemes in Figure \ref{subfig:N8SINR}, further increase in the dataset size may not contribute to the performance anymore.


After offline training, one model needs to be selected for deployment in real time. That is, we need to determine the epoch at which the model achieved the best combination of SSNR and SINR. Due to the aforementioned overfitting problem, the model state at the end of the training process may not correspond to the best sensing-communication tradeoff. Thus, a proper select-criterion should be followed to find the best student model throughout the training. A simple strategy is followed to find such model. Specifically, we look for the highest SSNR achieved within a certain range of epochs for which the achieved SINR is within a certain percentage of the maximum SINR performance. For instance, the maximum SINR of the blue curve in Figure \ref{subfig:N8SINR} is 1.5743. We search for the best SSNR within the epochs that correspond to an SINR of at least 1.4798 (i.e., 94\% of the maximum SINR). As such, out of ten epochs that satisfy the 94\% condition, the maximum SSNR is achieved at epoch number 222. The corresponding SINR and SSNR achieved by the model at epoch 222 are 1.4802 and 1.9338, respectively. The corresponding model parameters at this epoch can be selected for deployment in real-time.

\begin{figure}[!t]
    \centering
    
    \subfloat[\label{fig:modelStrat}]{\includegraphics[width=0.5\linewidth]{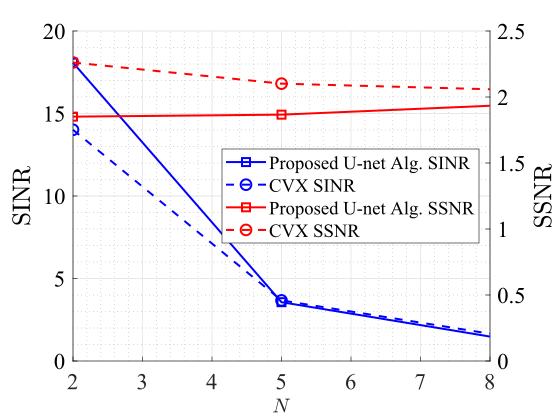}}
    \subfloat[\label{fig:runtime}]{\includegraphics[width=0.5\linewidth]{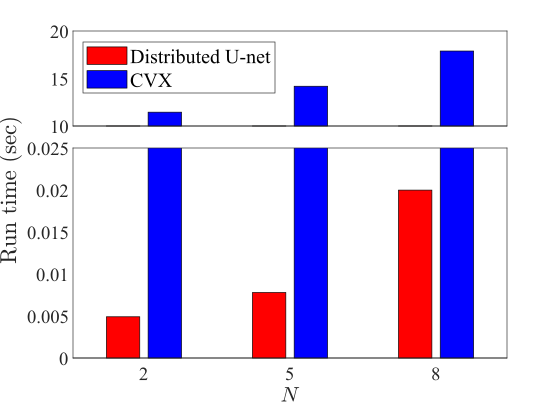}}

    \caption{(a) Selection strategy results at 94\% threshold percentage of SINR for different users. (b) Average run-time comparison, in seconds, of the proposed DL method using U-net and the CVX-based solution.}
    \label{fig}

\end{figure}

Applying the aforementioned strategy on the validation curves of $N=2,5$ and $8$, where the percentage threshold for the SINR is 94\%, yields the results shown in Figure \ref{fig:modelStrat}. While the DL method outperforms the CVX-based solution in terms of average SINR at low number of users, the corresponding average SSNR is lower than the CVX-based solution. As the number of users increases, the gap between the DL and the CVX-based outputs is reduced. The SINR achieved by the proposed DL method at higher number of users is slightly lower than the SINR ahchieved by the CVX-based algorithm.
    


\subsection{Time complexity}

While the performance comparison between the DL and the CVX-based algorithms varies depending on the architecture and the number of users, the running time of the proposed DL scheme is consistently much lower than the running time of the CVX-based algorithm.

To assist the running time of both methods, we consider the same settings as the previous section (i.e., 2 APs, 16 antennas per AP and different number of users). As such, the proposed DL scheme associates two U-nets to the system. We apply the two U-nets to 100 data points for a certain number of users, $N$. Every U-net reports its own running time. As the proposed method is distributed, we consider the maximum running time of the two reported time values rather than adding them. The CVX-based algorithm proposed in \cite{demirhan2023cell} consists of two steps; the bisection algorithm, where problem (\ref{eq:optPriority}) is solved repeatedly until the algorithm converges, and SDP to solve problem (\ref{eq:jsc2}). We report the total running time of both steps. We apply the CVX-based solution to the same 100 data points at a certain number of users , $N$. The numerical values of the described experiments  at $N=2,5$ and $8$ are shown in Figure \ref{fig:runtime}. For the 8 user case, we consider a filter size of $5\times 5$ throughout the U-net layers. Recall that the filter size of the U-net for $N=2$ and $N=5$ cases is $3\times 3$.


    

The time complexity of the proposed DL method is given by the general time complexity of CNNs, which is $\mathcal{O}\left(\sum_{d=1}^{D}(M_dN_d)f_d^2C_{d-1}C_d\right)$ \cite{wang2021fbunet}, where $D$ is the number of convolutional layers, $M_d$ and $N_d$ are the dimensions of the output size, $f_d$ is the filter size (assuming square filters) and $C_d$ is the number of channels. In Figure \ref{fig:runtime}, the DL run-time for 5 users is slightly longer than the run-time for 2 users due to the difference in only one dimension of the feature map sizes. On the other hand, the filter size used for the 8 user case is responsible for the large running time gap between the 8 user case and the other two user cases.


\section{Conclusion}
\label{sec:conc}

This work considers an unsupervised, distributed, teacher-student DL scheme to jointly optimize the beamformers in cell-free ISAC networks. The results show that the proposed method achieves a close performance to the conventional CVX solver's. While the CVX-based solution is centralized and time-consuming, the trained student models can be used in real-time situations, where each student is associated with one AP and can evaluate its own set of beamforming vectors without requiring CSI information from the other APs. The method was evaluated at different CNN architectures, and it was shown that U-net was the best performing architecture out of them.

Overfitting problem is a crucial aspect to the proposed method at higher number of UEs. It can be partially alleviated by increasing the model's filter size and increasing the dataset size. This work provides a benchmark for future works considering the same problem. More elaborate cooperation for the training process can be considered to further enhance the performance. It is also possible to consider DRL as an alternative approach for the cell-free ISAC beamforming problem.
\bibliographystyle{elsarticle-num} 
\bibliography{refs.bib}






\end{document}